\documentclass[
aps,%
12pt,%
final,%
notitlepage,%
oneside,%
onecolumn,%
nobibnotes,%
nofootinbib,%
superscriptaddress,%
noshowpacs]%
{revtex4}
\usepackage{graphicx}
\usepackage{color}
\usepackage{amsmath,amssymb}
\usepackage[T1]{fontenc}
\usepackage{array}
\usepackage{booktabs}          % modern table rules
\graphicspath{{figures/}}

\newcommand{\Msun}{M_\odot}
\newcommand{\cMpc}{\mathrm{cMpc}}
\newcommand{\cGpc}{\mathrm{cGpc}}
\newcommand{\MUV}{M_{\rm UV}}
\newcommand{\Mmol}{M_{\rm mol}}
\newcommand{\forg}{f_{\rm org}}
\newcommand{\Fbio}{F_{\rm bio}}
\newcommand{\Pabi}{p_{\rm abi}}
\newcommand{\Mcrit}{M_{\rm crit}}
\newcommand{\tabcell}[2]{\begin{minipage}[t]{#1}\raggedright\rule{0pt}{2.6ex}#2\rule[-1.1ex]{0pt}{0pt}\end{minipage}}
\definecolor{rulegray}{gray}{0.62}
\newcolumntype{I}{!{\color{rulegray}\vrule width 0.3pt}}   % thin inner column rule
\newcommand{\rowsep}{\noalign{{\color{rulegray}\hrule height 0.3pt}}}
\begin{document}

\noindent {\it Astronomy Reports, 2026, Vol. , No. }
\bigskip\bigskip  \hrule\smallskip\hrule
\vspace{35mm}

% Title footnote required by the special issue (to be completed with the meeting and editors):
% \footnote{Paper presented at ... Published by the recommendation of the special editors: ...}
\title{From Little Red Dots to Living Worlds: Following the First Prebiotic Molecules Through Cosmic Time}

\author{\bf \copyright $\:$  2026.
\quad \firstname{Yu}~\surname{Wang}}%
\email{yu.wang@icranet.org}
\affiliation{ICRA, Dipartimento di Fisica, Sapienza Universit\`a di Roma, Piazzale Aldo Moro 5, I-00185 Roma, Italy}%
\affiliation{ICRANet, Piazza della Repubblica 10, 65122 Pescara, Italy}%
\affiliation{INAF -- Osservatorio Astronomico d'Abruzzo, Via M. Maggini snc, I-64100 Teramo, Italy}%
\affiliation{Marcel Grossmann Center, Brickell Avenue 701, Miami, FL 33131, USA}%

\author{\bf \firstname{Remo}~\surname{Ruffini}}
\email{ruffini@icra.it}
\affiliation{ICRA, Dipartimento di Fisica, Sapienza Universit\`a di Roma, Piazzale Aldo Moro 5, I-00185 Roma, Italy}%
\affiliation{ICRANet, Piazza della Repubblica 10, 65122 Pescara, Italy}%
\affiliation{INAF, Viale del Parco Mellini 84, 00136 Rome, Italy}%

\author{\bf \firstname{Giorgio}~\surname{Sonnino}}
\email{giorgio.sonnino@ulb.be}
\affiliation{Universit\'e Libre de Bruxelles (ULB) \& International SOLVAY Institutes for Physics and Chemistry, Campus de la Plaine, P.C. 224, Build. NO, Bvd. du Triomphe, Brussels 1050, Belgium}%

\begin{abstract}

\centerline{\footnotesize Received: ;$\;$
Revised: ;$\;$ Accepted: .}\bigskip\bigskip\bigskip

Little Red Dots (LRDs), compact galaxies seen by the James Webb Space Telescope as early as 0.37~Gyr after the Big Bang ($z\approx12$), may have made the first prebiotic molecules in their cold, shielded cores.  We follow these molecules to the planets they could reach.  An LRD makes $\sim10^{55}$ prebiotic molecules in 10--30~Myr, and star formation, galaxy mergers, outflows, and dust grains carry them to planets.  For every LRD, about two planets that form from its gas stay habitable for more than a billion years, receive a steady supply of these molecules, and have the ultraviolet light, dry land, and phosphorus that prebiotic chemistry needs. Each of the two planets later forms long-lasting life with a probability $\Pabi=10^{-3}$--$1$.  The first such living worlds appear at $z\approx10$, and the entire observable Universe holds at least $9\times10^{10}\,\Pabi$ of them.  Today a galaxy like the Milky Way hosts $\sim0.7\,\Pabi$, but only $1\times10^{-8}\,\Pabi$ lie within 25~pc of the Sun, so it is no surprise that the Milky Way contains the Earth and that no second biosphere has been found near it.

\end{abstract}

\maketitle

\section{Introduction}
\label{sec:intro}

Every atom of carbon and nitrogen in a living cell was made in a star, and many of the molecules that first strung those atoms together may have been made in interstellar gas long before the Earth existed.  Laboratory and observational work has traced a path from simple interstellar species to complex organics~\cite{2009ARAandA..47..427H,2012AandARv..20...56C,2021PhR...893....1O,2022ApJS..259...30M}, and from small reactive molecules (nitriles, aldehydes, and phosphorus carriers) to amino acids and activated ribonucleotides~\cite{1953Sci...117..528M,2009Natur.459..239P,2015NatCh...7..301P}.  Observations can now address when and where the Universe first assembled the molecular ingredients of life, and whether those first ingredients reached a planet on which life began.

The James Webb Space Telescope (JWST) has found a possible site.  Little Red Dots (LRDs) are compact, reddened systems hosting accreting black holes~\cite{2024ApJ...963..129M,2024ApJ...968...38K,2025ApJ...986..126K}.  They differ from luminous active galactic nuclei (AGN).  Quasars, now classified and assigned redshifts in samples of hundreds of thousands of spectra~\cite{2022MNRAS.511.4490R,2024MNRAS.533.1976M}, are bright in the ultraviolet (UV) and illuminate the metal-enriched gas along their sightlines~\cite{2025ApJS..276...37L}, whereas LRDs are reddened and faint in X-rays.  The most distant known are at $z\approx12$, when the Universe was 0.37~Gyr old, a limit set by current selections rather than by the epoch of the first LRDs, and they are found down to $z\approx0.1$~\cite{2026ApJ...997..364L,2026arXiv260521574L}, and in JWST samples they are most common at $z\approx4.5$--$7.5$, 0.7--1.3~Gyr after the Big Bang.  Their black holes, of $10^{5}$--$10^{8}\,\Msun$~\cite{2026Natur.649..574R,2026Natur.653.1017J}, are active in compact, gas-rich environments.  Ruffini and Wang~\cite{2026ApJ..1000L..11R} argued that LRDs and the Milky Way's Central Molecular Zone (CMZ), the richest source of complex organic molecules in the Galaxy, which include nitrile precursors of nucleobases~\cite{1996ARAandA..34..645M,2006AandA...455..971R,2008ApJ...672..352R,2018MNRAS.478.2962Z}, are ``islands of electromagnetic tranquility.''  Both host black holes of about a million solar masses inside dense, cold, dusty molecular reservoirs, and both are electromagnetically quiet: Sgr~A$^*$ radiates below $10^{-9}$ of its Eddington luminosity, and LRDs show weak or undetected X-ray emission, no radio jets, and little variability.  Without the X-ray-dominated regions that dissociate molecules over hundreds of parsecs around quasars, such cores can build up complex organic chemistry.  Ruffini and Wang~\cite{2026ApJ..1000L..11R} proposed that LRDs formed the first prebiotic molecules, and that galactic winds and mergers then carried them to later generations of galaxies and planetary systems.  If so, prebiotic molecules were already being made a few hundred Myr after the Big Bang.  Here we test this proposal quantitatively.

\begin{figure}
\centering
\includegraphics[width=\textwidth]{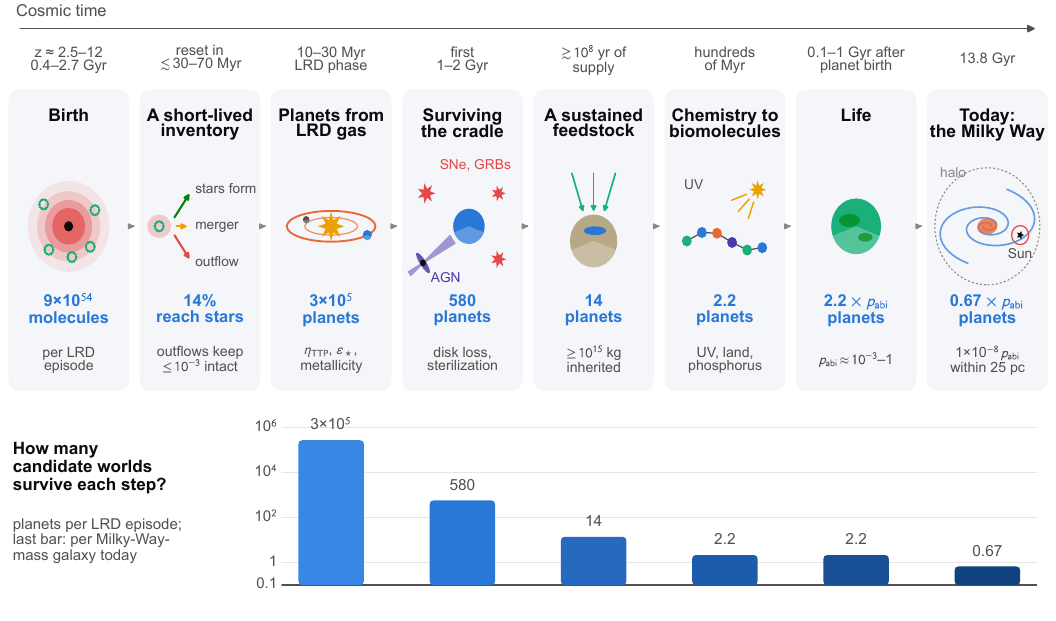}
\caption{The journey from Little Red Dots to living worlds.  LRDs are detected from $z\approx12$, 0.37~Gyr after the Big Bang, and the JWST sample used here spans $z\approx2.5$--$12$ (0.4--2.7~Gyr); eight stages are ordered by cosmic time (top), and each card shows what happens and the median number that survives per LRD episode.  Molecules made in an LRD are chemically short-lived outside dense ices; they reach planets in gas that forms stars during the LRD phase, in gas mixed in or thrown out by mergers, in gas returned by outflows, and on dust grains.  The bars at the bottom show how the number of temperate rocky planets formed from LRD gas falls at each subsequent step; the last bar is the number around stars still shining today in a Milky-Way-mass galaxy, per unit abiogenesis probability $\Pabi$.}
\label{fig:schematic}
\end{figure}

Many steps separate a molecule made in an LRD from a planet that sustains life, and each has been studied in a different field: how much cold gas LRDs contain and how organic it is; how long the molecules survive and where they go; how many temperate rocky planets form, and whether they survive their violent birthplace and remain habitable; whether a planet receives a sustained supply of feedstock; whether the chemistry that turns feedstock into biomolecules can run there; and whether life then emerges.  We follow the molecules through all of these steps in the order they happen, adopting at each the values supported by published observations, experiments, and simulations (Fig.~\ref{fig:schematic}).  The number of candidate worlds decreases at each step.  The result is a conditional count of living worlds fed by LRD chemistry, given per LRD episode, per cosmic epoch from $z\approx12$ onward, and per galaxy.  We compare it with the one available observation, a single known biosphere and no detection of a second.

Section~\ref{sec:journey} sets out the chain.  Sections~\ref{sec:birth}--\ref{sec:life} follow it in time, from the birth of the molecules to the emergence of life.  Section~\ref{sec:epochs} gives the resulting counts across cosmic time.  Section~\ref{sec:milkyway} applies them to the present-day Milky Way and compares them with observations.  Section~\ref{sec:discussion} discusses what controls the answer and how to test it.  We adopt a flat Planck 2018 cosmology~\cite{2020AandA...641A...6P} and the stellar initial mass function (IMF) of Chabrier~\cite{2003PASP..115..763C}.

\section{The chain of events}
\label{sec:journey}

We follow one \emph{LRD episode}: the $\sim10$--$30$~Myr during which a galaxy is observed as an LRD, together with the stars and planets formed from its gas.  The number of planets that can sustain life fed by LRD-made molecules is the product
\begin{equation}
N_{\rm life}=N_{\rm TTP}\;f_{\rm env}\;f_{\rm sust}\;f_{\rm pass}\;f_{\rm chem}\;\Pabi ,
\label{eq:chain}
\end{equation}
where $N_{\rm TTP}$ is the number of temperate terrestrial planets (TTPs) formed from LRD gas while its organics are intact, $f_{\rm env}$ the fraction whose disks and orbits survive the compact nuclear environment, $f_{\rm sust}$ the fraction that remain habitable long enough for life to arise and persist, $f_{\rm pass}$ the fraction receiving a sustained supply of inherited prebiotic feedstock, $f_{\rm chem}$ the fraction whose surface allows the chemistry that turns feedstock into biomolecules, and $\Pabi$ the probability that long-lasting life then emerges.  Table~\ref{tab:ledger} lists every factor, the range we adopt, the number that survives it, and the section that gives its literature basis.

Each range is sampled log-uniformly in $2\times10^{4}$ Monte Carlo draws, because most span a factor of several or more with no preferred value inside them, and we quote medians with 16th--84th percentiles.  The organic content of a planet's natal core and its delivery history vary from planet to planet, and we treat them as distributions over planets within each draw.  The abiogenesis probability $\Pabi$ is the probability that a planet which stays habitable for long enough, receives feedstock, and can run the intervening chemistry develops self-replicating, evolving life; the life is long-lasting because we require the planet to stay habitable for at least a gigayear after it emerges (Section~\ref{sec:planets}).  It has not been measured.  If the Earth is typical, its early biosphere implies $\Pabi>0.13$ at 95\% confidence~\cite{2002AsBio...2..293L}, and a Bayesian treatment of the timing favours rapid abiogenesis~\cite{2020PNAS..11711995K}.  Because the Earth is a single example subject to selection, however, the posterior depends on the prior, and the data give no lower bound~\cite{2012PNAS..109..395S}.  We therefore report counts per unit $\Pabi$ and evaluate them over the range $\Pabi=10^{-3}$--$1$.

\begin{table}
\renewcommand{\baselinestretch}{1}\selectfont
\centering
\caption{The ledger of the journey, per LRD episode}
\label{tab:ledger}
\footnotesize\setlength{\tabcolsep}{4pt}
\begin{tabular}{lIlIlIlIlIl}
\toprule
\tabcell{0.14\textwidth}{Stage, time} & \tabcell{0.19\textwidth}{Condition} & \tabcell{0.245\textwidth}{Adopted values} & \tabcell{0.08\textwidth}{Step factor} & \tabcell{0.19\textwidth}{Number, median (16--84\%)} & \tabcell{0.04\textwidth}{Sec.} \\
\midrule
\tabcell{0.14\textwidth}{\textit{Prebiotic}\newline \textit{molecules}} & & & & & \\
\tabcell{0.14\textwidth}{Birth\newline $z\approx2.5$--$12$\newline (this sample)} & \tabcell{0.19\textwidth}{Synthesis in cold host gas} & \tabcell{0.245\textwidth}{$\Mmol^0=3\times10^{7}$--$3\times10^{8}\,\Msun$\newline $Z_{\rm LRD}=0.03$--$0.2\,Z_\odot$\newline $\forg=f_{{\rm org},\odot}(Z_{\rm LRD}/Z_\odot)\,\delta$} & \tabcell{0.08\textwidth}{--} & \tabcell{0.19\textwidth}{$9\times10^{54}$\newline ($2\times10^{54}$--$5\times10^{55}$)} & \tabcell{0.04\textwidth}{\ref{sec:birth}} \\
\rowsep
\tabcell{0.14\textwidth}{Star formation\newline 10--30~Myr} & \tabcell{0.19\textwidth}{Gas forms stars before its organics are destroyed} & \tabcell{0.245\textwidth}{$\epsilon_\star=0.05$--$0.4$, $t_{\rm LRD}=10$--$30$~Myr\newline outflow survival $\leq10^{-3}$ (gas, ices)} & \tabcell{0.08\textwidth}{0.14} & \tabcell{0.19\textwidth}{$1.3\times10^{54}$\newline ($2\times10^{53}$--$8\times10^{54}$)} & \tabcell{0.04\textwidth}{\ref{sec:travel}} \\
\midrule
\tabcell{0.14\textwidth}{\textit{Temperate}\newline \textit{rocky planets}} & & & & & \\
\tabcell{0.14\textwidth}{Planets form\newline LRD phase} & \tabcell{0.19\textwidth}{Rocky planets form in the habitable zone} & \tabcell{0.245\textwidth}{$\eta_{\rm FGK}=0.07$--$0.13\,\Msun^{-1}$\newline $\eta_M=0.2$--$0.45\,\Msun^{-1}$, $f_M=0.01$--$0.5$\newline $Z_c=0.03$--$0.3\,Z_\odot$} & \tabcell{0.08\textwidth}{--} & \tabcell{0.19\textwidth}{$2.8\times10^{5}$\newline ($6\times10^{4}$--$9.5\times10^{5}$)} & \tabcell{0.04\textwidth}{\ref{sec:planets}} \\
\rowsep
\tabcell{0.14\textwidth}{Nucleus\newline first Myr} & \tabcell{0.19\textwidth}{Disks, ice line, and orbits survive} & \tabcell{0.245\textwidth}{$f_{\rm env}=0.1$--$0.5$} & \tabcell{0.08\textwidth}{0.22} & \tabcell{0.19\textwidth}{$6.1\times10^{4}$\newline ($1.3\times10^{4}$--$2.3\times10^{5}$)} & \tabcell{0.04\textwidth}{\ref{sec:planets}} \\
\rowsep
\tabcell{0.14\textwidth}{Habitability\newline next 1--2~Gyr} & \tabcell{0.19\textwidth}{No sterilization for $t_{\rm abi}+1$~Gyr} & \tabcell{0.245\textwidth}{$f_{\rm sust}=10^{-3}$--$0.1$\newline $t_{\rm abi}=0.1$--$1$~Gyr} & \tabcell{0.08\textwidth}{0.010} & \tabcell{0.19\textwidth}{580 (75--4300)} & \tabcell{0.04\textwidth}{\ref{sec:planets}} \\
\rowsep
\tabcell{0.14\textwidth}{Feedstock\newline $\gtrsim10^{8}$~yr} & \tabcell{0.19\textwidth}{Sustained inherited supply, $D\geq1$} & \tabcell{0.245\textwidth}{$\epsilon_{\rm planet}=10^{-9}$--$10^{-7}$\newline $\Mcrit=10^{15}$--$10^{16}$~kg} & \tabcell{0.08\textwidth}{0.033\newline (0.024)} & \tabcell{0.19\textwidth}{14 (0.3--200)} & \tabcell{0.04\textwidth}{\ref{sec:life}} \\
\rowsep
\tabcell{0.14\textwidth}{Chemistry\newline $\sim10^{8}$~yr} & \tabcell{0.19\textwidth}{UV light, exposed land, reactive phosphorus} & \tabcell{0.245\textwidth}{$f_{\rm chem}\approx0.2$ (FGK)\newline $f_{\rm chem}\approx10^{-3}$ (M)} & \tabcell{0.08\textwidth}{0.15} & \tabcell{0.19\textwidth}{2.2 (0.04--31)} & \tabcell{0.04\textwidth}{\ref{sec:life}} \\
\rowsep
\tabcell{0.14\textwidth}{Life\newline $+0.1$--$1$~Gyr} & \tabcell{0.19\textwidth}{Self-replication and evolution} & \tabcell{0.245\textwidth}{$\Pabi=10^{-3}$--$1$} & \tabcell{0.08\textwidth}{$\Pabi$} & \tabcell{0.19\textwidth}{$2.2\,\Pabi$\newline (0.04--31)} & \tabcell{0.04\textwidth}{\ref{sec:life}} \\
\rowsep
\tabcell{0.14\textwidth}{Host star\newline today} & \tabcell{0.19\textwidth}{Still on the main sequence} & \tabcell{0.245\textwidth}{turnoff mass $\approx0.8\,\Msun$} & \tabcell{0.08\textwidth}{0.44} & \tabcell{0.19\textwidth}{$0.95\,\Pabi$ (0.02--14)} & \tabcell{0.04\textwidth}{\ref{sec:milkyway}} \\
\rowsep
\tabcell{0.14\textwidth}{Milky Way\newline today} & \tabcell{0.19\textwidth}{Episodes per galaxy of $(5$--$6)\times10^{10}\,\Msun$} & \tabcell{0.245\textwidth}{0.7 (0.5--1.1)} & \tabcell{0.08\textwidth}{0.72} & \tabcell{0.19\textwidth}{$0.67\,\Pabi$ (0.014--10)} & \tabcell{0.04\textwidth}{\ref{sec:milkyway}} \\
\rowsep
\tabcell{0.14\textwidth}{Near the Sun\newline today} & \tabcell{0.19\textwidth}{Within 25~pc of the Sun} & \tabcell{0.245\textwidth}{$2\times10^{-9}$--$1.2\times10^{-7}$ of the Galaxy} & \tabcell{0.08\textwidth}{$2\times10^{-8}$} & \tabcell{0.19\textwidth}{$1.0\times10^{-8}\,\Pabi$\newline ($2\times10^{-10}$--$2\times10^{-7}$)} & \tabcell{0.04\textwidth}{\ref{sec:milkyway}} \\
\bottomrule
\end{tabular}

\medskip
\begin{flushleft}\footnotesize Medians (16th--84th percentiles) of $2\times10^4$ Monte Carlo draws with log-uniform ranges, per LRD episode except in the last two rows.  Step factor: median over draws of the fraction passing the step; in parentheses, the ratio of successive medians where it differs, owing to correlations between draws.  Sources: listed section and Appendix~\ref{sec:param_appendix}.\end{flushleft}
\end{table}
\section{Birth: prebiotic molecules in Little Red Dots}
\label{sec:birth}

\subsection{How many LRDs}

We use all 341 LRDs of Kocevski et~al.~\cite{2025ApJ...986..126K}, found in six JWST fields covering $588\,\mathrm{arcmin}^2$ at $z=2.44$--$11.92$.  Each source is weighted by the inverse of the comoving volume over which it would have passed the catalog selection, $\mathrm{S/N(F444W)}>12$, in any of the six fields, following the coherent accessible-volume ($1/V_a$) estimator~\cite{1968ApJ...151..393S,1980ApJ...235..694A}:
\begin{equation}
n_{\rm LRD}(z_b)=\sum_{i\in b}\frac{1}{V_{a,i}},\quad
V_{a,i}=\sum_{f}\Omega_f\!\int_{z_{b,1}}^{\min(z_{b,2},z_{{\rm max},i,f})}\!\frac{{\rm d}V}{{\rm d}z\,{\rm d}\Omega}{\rm d}z ,
\label{eq:nlrd}
\end{equation}
in bins of $\Delta z=0.5$, where $z_{{\rm max},i,f}$ is the redshift at which source $i$ would fall below the selection limit of field $f$ (Appendix~\ref{sec:density_appendix}).  The UV luminosity function this gives agrees with the published one of Kocevski et~al.~\cite{2025ApJ...986..126K} within 0.25~dex at $\MUV\le-18$ (Appendix~\ref{sec:density_appendix}).  At $\MUV=-17$ our F444W-based volumes leave the sample $\sim0.5$~dex less complete; that bin holds 18\% of the LRDs but only 2.5\% of their UV-weighted gas, and we absorb the residual into a selection factor $C_{\rm sel}=1$--$1.6$.  LRDs are already present at $z\approx12$, with $\sim10^{-5}\,\cMpc^{-3}$ at $z\approx10$--$12$, when the Universe was 0.37--0.47~Gyr old.  This is the highest redshift of the sample, not necessarily the epoch of the first LRDs: LRDs are identified by their Balmer break, at a rest wavelength of 3646~\AA, and red rest-frame optical continuum, and at $z\approx12$ the break reaches the red edge of the reddest NIRCam filter, F444W, at $\approx5\,\mu$m, so earlier LRDs would be missed by current selections.  The abundance peaks at $(4$--$9)\times10^{-5}\,\cMpc^{-3}$ over $z\simeq4.5$--$7.5$ and, in this sample, declines below $z\approx4.5$ (Appendix~\ref{sec:density_appendix}; see Section~\ref{sec:epochs} for lower redshifts).

\subsection{How much cold gas}

Prebiotic molecules form in cold molecular gas, and the stars and planets of an LRD form from the same gas.  We therefore need the total mass of this gas.  It is almost entirely H$_2$ and helium; the prebiotic molecules are a small fraction of it, derived in Section~\ref{sec:organic}.

This gas is not the dense envelope that produces the LRD spectra in black-hole-star models, which is warm ($5000$--$10^4$~K), partly ionized, nearly dust-free, and holds $\lesssim10^{7}\,\Msun$~\cite{2025ApJ...980L..27I,2025AandA...701A.168D,2026Natur.656..329N,2025MNRAS.544.3407K}.  It is the interstellar medium of the host on 30--200~pc scales, the counterpart of the Milky Way's CMZ, which holds $\sim10^{7}\,\Msun$ of molecular gas within 100--250~pc, with dust at 15--40~K and gas at 50--200~K~\cite{2026ApJ..1000L..11R}.

No cold gas has yet been measured directly in an LRD.  Ruffini and Wang~\cite{2026ApJ..1000L..11R} inferred it from the stellar masses of LRD hosts and the gas-to-stellar mass ratios of high-redshift galaxies: $10^{6}$--$10^{9}\,\Msun$ of molecular gas, with dust at $\lesssim50$~K, about half the dust temperature around an ordinary AGN because the accretion luminosity of LRDs is low, and gas at 50--200~K.  The corresponding dust masses, $10^{3}$--$10^{6}\,\Msun$, agree with ALMA dust-continuum and [C\,{\sc ii}] observations of LRDs~\cite{2025ApJ...991L..10S,2025AandA...700A.231X,2026ApJ...997..218A}, whose stacked continuum limits the dust to $\lesssim10^{6}\,\Msun$~\cite{2025ApJ...990L..61C}.

We adopt this range and assign it to each LRD through its UV luminosity, which traces the star formation of its host~\cite{2014ARAandA..52..415M}: $\Mmol=\min(\Mmol^{0}10^{-0.4(\MUV+20)},10^{9}\,\Msun)$, with $\Mmol^{0}=3\times10^{7}$--$3\times10^{8}\,\Msun$ at $\MUV=-20$ and the cap at the upper end of the inferred range.  For the sample, with $\MUV\approx-21.3$ to $-16.3$ for 98\% of the sources, this gives $10^{6}$--$10^{9}\,\Msun$ per LRD; even for the largest $\Mmol^{0}$ only 4 of the 341 LRDs reach the cap.

We use the gas mass rather than the dust mass because the abundances of complex organics are measured relative to H$_2$ (Appendix~\ref{sec:param_appendix}) and because stars form from the gas.  The dust mass is used only as a check, and the smaller grain surface at low metallicity enters through the organic fraction of Section~\ref{sec:organic}, which turns this gas into $\approx0.2\,\Msun$ of prebiotic organics per episode.

\subsection{How organic, and when the elements are there}
\label{sec:organic}

The organic content of the gas depends on its metallicity.  Direct-method measurements of 24 LRD hosts at $z\approx2.3$--$7$ give a mean of $0.08\,Z_\odot$ with a spread of $\sim0.6$~dex and no trend with redshift~\cite{2026arXiv260631515N}, and the dense gas around the central engine of LRDs has only $\sim1\%$ of the solar metallicity~\cite{2026arXiv260909271K}.  No LRD metallicity has been measured above $z\approx7$, but galaxies at $z\approx9$--$14$ have similar values, $0.05$--$0.2\,Z_\odot$~\cite{2023AandA...677A..88B,2024AandA...684A..75C,2025NatAs...9..155Z,2025ApJ...988...19S,2026ApJ..1001...38T,2026AandA...708A.203P}.  We adopt $Z_{\rm LRD}=0.03$--$0.2\,Z_\odot$ at all redshifts; the results for LRDs above $z\approx7$ are therefore conditional on this assumption.

In the CMZ cloud G+0.693$-$0.027, the complex organics that are prebiotic precursors make up a mass fraction $f_{{\rm org},\odot}=5\times10^{-7}$--$1.2\times10^{-6}$ of the molecular gas~\cite{2008ApJ...672..352R,2018MNRAS.478.2962Z,2020AsBio..20.1048J,2021PNAS..11801314R}.  They include nitriles, isocyanic acid, formamide, imines, glycolaldehyde, ethylene glycol, and the more recently detected urea, hydroxylamine, and ethanolamine.  In Magellanic Cloud hot cores, complex organics scale roughly with metallicity in some sources, such as N~113 and N~105~\cite{2018ApJ...853L..19S,2022ApJ...931..102S}, while organic-poor cores such as ST11 fall below that scaling by at least 1--3~dex~\cite{2016ApJ...827...72S,2020ApJ...891..164S}, and the scatter is similarly large in the Small Magellanic Cloud~\cite{2023ApJ...946L..41S}.  A lower dust-to-gas ratio also reduces the grain surface on which the ices that seed prebiotic molecules form, so that prebiotic chemistry proceeds at low metallicity but with reduced efficiency and large variations from source to source~\cite{2026ApJ..1000L..11R}.  We write $\forg=f_{{\rm org},\odot}(Z_{\rm LRD}/Z_\odot)\,\delta$, with an extra depletion $\delta=10^{-2}$--$1$ that varies from core to core: $\delta=1$ corresponds to the metallicity-scaled cores and $10^{-2}$ to the organic-poor ones.  Extending $\delta$ to $10^{-3}$ would lower the final count by one third.  The median is $\forg=6\times10^{-9}$.

The elements are available early.  JWST detects nitrogen at $z=12$--$14$, often with a nitrogen-to-oxygen ratio above solar~\cite{2023MNRAS.523.3516C,2024ApJ...966...92S,2024ApJ...972..143C,2026OJAp....956033N}, produced by Wolf--Rayet winds within $\sim10$~Myr of a starburst or by asymptotic-giant-branch stars after 30--40~Myr~\cite{2026MNRAS.tmp.1597R}, although such enhancements are detected in only a minority of galaxies~\cite{2025ApJ...994L..29Z}.  Carbon is present at $z=12.5$~\cite{2024AandA...689A.152D}, and phosphorus tracks iron in metal-poor stars, as expected for a massive-star origin~\cite{2014ApJ...797...69R,2014ApJ...796L..24J}.  The availability of carbon, hydrogen, nitrogen, oxygen, phosphorus, and sulfur (CHNOPS) is set by the overall metallicity.  Ruffini and Wang~\cite{2026ApJ..1000L..11R} noted that the first galaxies form at $z\sim20$, when the cosmic microwave background has cooled to $\sim50$~K, the upper dust temperature for efficient grain-surface chemistry.  We include only a prompt nitrogen timescale of 10--40~Myr after the first star formation at this epoch (Appendix~\ref{sec:gates_appendix}), which makes the CHNOPS factor $\Fbio\approx1$ at every LRD epoch.  A single LRD episode then assembles $\approx0.2\,\Msun$ of prebiotic organics, or $9\times10^{54}$ molecules of 30 atomic mass units, and the prebiotic organic density is $\rho_{\rm org}\approx(1.2$--$2.6)\times10^{4}\,\Msun\,\cGpc^{-3}$ at $z=4.5$--$7.5$ and $\sim(0.6$--$4)\times10^{3}\,\Msun\,\cGpc^{-3}$ at $z\approx10$--$12$.

\section{A short-lived inventory: where the molecules go}
\label{sec:travel}

\begin{figure}
\centering
\includegraphics[width=\textwidth]{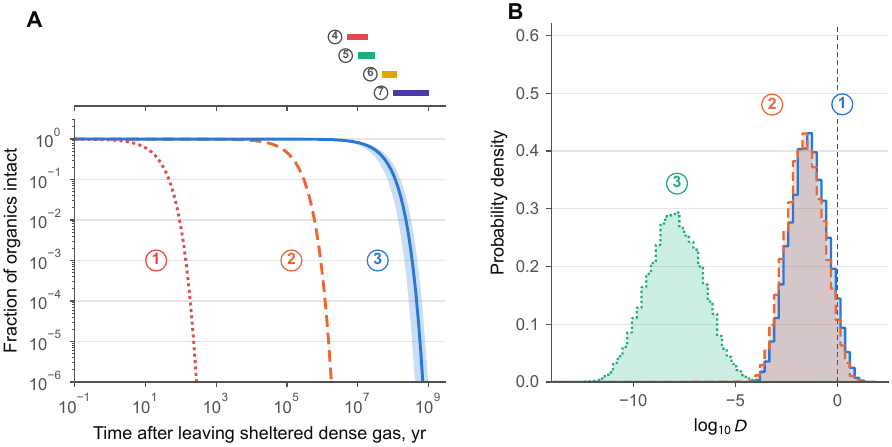}
\caption{A short-lived inventory.  (A) Fraction of prebiotic organics still intact after they leave sheltered dense gas: 1, unshielded gas (UV photodissociation, $\sim20$~yr); 2, shielded gas (cosmic rays, $\sim0.1$~Myr); 3, ices in dense clouds (30--70~Myr, band).  The strip above marks the timescales of the processes that move LRD gas: 4, destruction of carrier clouds in winds; 5, star formation during the LRD phase; 6, merger coalescence; 7, outflow recycling.  (B) Distribution over Monte Carlo draws of the median dose ratio $D$ (Equation~\ref{eq:dose}) for the $z=7.0$--$7.5$ bin: 1, planets formed from LRD gas (the median planet reaches $D\geq1$ in 6\% of draws); 2, after merger mixing (4\%); 3, organics recycled through outflows (none).  The dashed line marks $D=1$, a sustained supply.}
\label{fig:travel}
\end{figure}

Ruffini and Wang~\cite{2026ApJ..1000L..11R} suggested that LRD-made organics could be carried by galactic winds, incorporated into protoplanetary disks, and spread by mergers.  The abundances of interstellar organics are set by a balance between formation and destruction.  Outside dense gas the interstellar UV field photodissociates methanol, acetaldehyde, and formamide within 12--23~yr, and in fully shielded gas cosmic rays destroy them within 0.1--0.2~Myr~\cite{2017AandA...602A.105H}; organics locked in the ices of dense clouds last longer, 30--70~Myr~\cite{2001ApJ...550L..95E}.  Molecules that leave the sheltered LRD gas are therefore replaced by whatever the new environment produces within at most a few tens of Myr (Fig.~\ref{fig:travel}A).

\emph{Outflows.}  Compact starbursts and AGN hosts expel molecular gas, including dense-gas tracers~\cite{2017ApJ...835..265W}, at one to several times their star-formation rate~\cite{2013Natur.499..450B,2019ApJ...881...43K,2019MNRAS.483.4586F}.  The gas launched in one episode is this mass-loading factor times the gas turned into stars; for a loading of unity and the $\epsilon_\star$ adopted below, it is a fraction $\epsilon_{\rm launch}=0.05$--$0.4$ of the cold gas.  Larger loadings do not change the result, because even lossless transport leaves the dose below unity (see below).  Of the ejected gas, 50--90\% is re-accreted by the same galaxy after 0.1--1~Gyr, landing well outside the nucleus, and less than 10\% is transferred to other galaxies~\cite{2016ApJ...824...57C,2017MNRAS.470.4698A,2019MNRAS.490.4786G}.  Most of the molecules are destroyed on the way.  Carrier clouds in galactic winds are shredded or re-condensed within a few to 20~Myr~\cite{2015ApJ...805..158S,2018MNRAS.480L.111G,2023MNRAS.524.1258N}; dust grains larger than $\sim0.01\,\mu$m largely survive galactic winds, while smaller grains are destroyed in the hot phase~\cite{2024ApJ...974...81R}; and in NGC~253, dense-gas tracers are seen in an outflow about 1~Myr old~\cite{2017ApJ...835..265W}, whereas complex organics are observed only in its central molecular zone~\cite{2025AandA...698A.261B}.  Molecules in gas and volatile ices survive the recycling time only if both their carrier cloud (a few to 20~Myr) and their ices (30--70~Myr) outlast it; for $t_{\rm rec}=0.1$--$1$~Gyr the factors $\exp(-t_{\rm rec}/\tau)$ span many orders of magnitude, and we adopt an intact fraction $f_{\rm surv}=10^{-6}$--$10^{-3}$, a choice on which the result does not depend.  The survivors are then diluted into all the gas that forms the galaxy's stars during recycling, and their dose on a new planet is $D\approx1\times10^{-8}$ of what prebiotic chemistry needs (Section~\ref{sec:life}; Fig.~\ref{fig:travel}B).  A refractory fraction carried on dust grains survives longer (Appendix~\ref{sec:earth_appendix}), but even an intact fraction of 0.1 gives $D\lesssim10^{-4}$, and lossless transport gives $D\approx4\times10^{-3}$.  Outflows redistribute the carbon, nitrogen, and oxygen of the LRD together with a fraction $\epsilon_{\rm launch}f_{\rm surv}f_{\rm rec}$ of its molecules.

\emph{Mergers.}  At $z\sim5$--$8$, galaxies of $\sim10^{8}\,\Msun$ undergo 4--12 mergers per Gyr with mass ratios above 1:10~\cite{2025MNRAS.540..774D,2025MNRAS.540.2146P,2026arXiv260218068C}, and many LRDs have companions within a few kiloparsecs~\cite{2024arXiv241214246T,2026ApJ..1002L...4B,2026arXiv260527903B}, although on megaparsec scales they sit in underdense regions~\cite{2025ApJ...989L..50C}.  With the merger timescales of Conselice et~al.~\cite{2022ApJ...940..168C}, as adopted by Duan et~al.~\cite{2025MNRAS.540..774D}, close pairs at $z\approx4.5$--$7.5$ coalesce within 0.05--0.13~Gyr for mass ratios from 1:1 to 1:10.  A merger mixes LRD gas with its partner's before the LRD organics are reset only if it occurs within the lifetime of the organics, which happens with probability $\approx0.36$ (Appendix~\ref{sec:param_appendix}).  Mixing lowers the organic fraction but raises the number of planets forming from LRD-bearing gas by nearly the same factor, and mergers change the final count by 7\%.  Later mergers act on gas whose organics have already been reset, or relocate stars and planets that have already formed (Section~\ref{sec:milkyway}).

\emph{Mergers that displace LRD gas.}  A merger can also carry LRD gas away from the nucleus.  In gas-rich mergers most of the nuclear gas loses angular momentum to gravitational torques and falls further in, where it is consumed in a central starburst~\cite{1996ApJ...471..115B,2009ApJ...691.1168H}; this is the mixing described above.  The tidal tails and tidal dwarf galaxies formed in mergers consist mainly of material torn from the outer disks, and their molecular gas forms in situ from the expelled atomic gas~\cite{2001AandA...378...51B,2013LNP...861..327D}.  Molecular gas from the inner regions can nevertheless be thrown out: in the minor merger NGC~4194 at least $8.5\times10^{7}\,\Msun$ of molecular gas, 4\% or more of that of the system, lies in a tidal tail~\cite{2001AandA...372L..29A}.  When the LRD is the smaller partner, as expected for hosts of $10^{7.5}$--$10^{8.5}\,\Msun$, its compact core survives the accretion and sinks to the centre of the larger galaxy~\cite{2007MNRAS.374.1227B,2023ApJ...955L..12B}, while gas stripped along its orbit is spread over the outskirts.  Continuous stripping of dense nuclear gas into molecular tails is observed in cluster galaxies~\cite{2017ApJ...839..114J}, although LRDs mostly lie in underdense regions~\cite{2025ApJ...989L..50C}.  The displaced gas reaches the outskirts over the 0.05--0.13~Gyr of the merger.  Gas-phase molecules in the entrained clouds are dissociated within a few to 10~Myr, as in the Milky Way's nuclear wind~\cite{2023MNRAS.524.1258N}, and the fraction of ice-locked organics still intact on arrival is $\exp(-t/\tau_{\rm ice})\approx0.01$--$0.5$ for $\tau_{\rm ice}=30$--$70$~Myr.  The displaced gas then forms stars and planets in the tidal features, which receive the surviving LRD organics together with those formed in situ.  We describe this route but do not add it to the count, because the fraction of nuclear gas that mergers expel is constrained by a single system.

\emph{Staying home.}  The most direct route is through the gas that forms stars during the LRD phase.  In nascent-starburst models this phase lasts $\sim15$~Myr and ends when stellar feedback expels the nuclear gas~\cite{2026ApJ..1000...90I}; related models give 20--30~Myr~\cite{2026ApJ...998L...4S,2025MNRAS.544.3407K}.  This exceeds the $\gtrsim10^{5}$--$10^{6}$~yr for which the tranquil configuration must persist to build up prebiotic molecules~\cite{2026ApJ..1000L..11R}.  We take $t_{\rm LRD}=10$--$30$~Myr and a fraction $\epsilon_\star=0.05$--$0.4$ of the reservoir turned into stars before expulsion, spanning the efficiency expected for $10^{7}$--$10^{8}\,\Msun$ within 100~pc from the surface-density dependence of Grudić et~al.~\cite{2018MNRAS.475.3511G}, the low efficiency of the CMZ~\cite{2013MNRAS.429..987L}, and cloud-scale values~\cite{2020MNRAS.493.2872C}.  Turbulence mixes gas across 100~pc within 5--10~Myr~\cite{2012ApJ...758...48Y,2018MNRAS.475.2236K}, so this star-forming gas carries the host's organic fraction.  Of the $9\times10^{54}$ molecules made per episode, $1.3\times10^{54}$ (14\%) are still intact when their gas forms stars.

\section{Planets born in a hostile cradle}
\label{sec:planets}

\emph{How many planets.}  Habitable-zone rocky planets orbit 0.37--0.60 of FGK stars~\cite{2021AJ....161...36B} and 0.16--0.33 of M dwarfs~\cite{2015ApJ...807...45D,2020MNRAS.498.2249H}.  With 0.19--0.22 FGK stars and 1.2--1.4 M dwarfs per solar mass formed, this gives $\eta_{\rm FGK}=0.07$--$0.13$ and $\eta_M=0.2$--$0.45$ temperate rocky planets per solar mass.  M-dwarf planets are exposed to an intense early activity that may strip their atmospheres, and we count a fraction $f_M=0.01$--$0.5$ of them.  The upper end lets the planets of early M dwarfs keep their atmospheres; the lower end reflects that around mid-to-late M dwarfs only the largest rocky planets do~\cite{2025ApJ...986L...3P} and that most M-dwarf planets are unlikely to host life~\cite{2016PhR...663....1S,2018JCAP...05..020L}.  Terrestrial planets need a minimum metallicity, about $0.03\,Z_\odot$ at 1~AU~\cite{2012ApJ...751...81J}; small planets depend only weakly on metallicity above $[{\rm Fe/H}]\approx-0.4$~\cite{2018AJ....155...89P}, while close-in super-Earths become rare near $0.2$--$0.3\,Z_\odot$~\cite{2024AJ....168..128B}.  We apply a gate $F_Z=[1+(Z_c/Z_{\rm LRD})^2]^{-1}$ with $Z_c=0.03$--$0.3\,Z_\odot$; for the metal-poor LRD gas its median is 0.40.  Each LRD episode then forms $2.8\times10^{5}$ temperate rocky planets from its organic-bearing gas.

\emph{Surviving the cradle.}  The nuclear environment in which these planets form is hostile.  In the Milky Way's CMZ, external photoevaporation by nearby massive stars destroys 90\% of protoplanetary disks within 1~Myr~\cite{2020MNRAS.491..903W}; planet-forming disks survive only at low stellar surface density~\cite{2021MNRAS.501.1782C}; above a stellar surface density of $\sim6\times10^{3}\,\Msun\,{\rm pc^{-2}}$, for the Galactic dust-to-gas ratio, disks have no ice line and planets form dry~\cite{2013MNRAS.431...63T}; and in dense stellar systems close encounters destroy planetary systems, the orbital lifetime of a planet being $\tau\approx3\times10^{8}\,{\rm yr}\,(v/20\,{\rm km\,s^{-1}})(10^{6}\,{\rm pc^{-3}}/n)(0.1\,{\rm au}/a)(0.5\,\Msun/m_t)$ for a stellar density $n$, velocity dispersion $v$, orbital radius $a$, and combined mass $m_t$ of the encounter~\cite{2016ApJ...827...54D}.  A planet at 1~au around a Sun-like star ($m_t\approx1.5\,\Msun$) then survives a gigayear only where $n\lesssim10^{4}\,{\rm pc^{-3}}\,(v/20\,{\rm km\,s^{-1}})$, so habitable-zone orbits are stable only in the less crowded regions.  We adopt a survival fraction $f_{\rm env}=0.1$--$0.5$: the lower end is the 10\% of CMZ disks that survive their first Myr, and the upper end allows up to half of the stars to form in less crowded parts of the host, below the $\sim100$~stars~pc$^{-2}$ at which planet-forming disks survive~\cite{2021MNRAS.501.1782C}.  This leaves $6.1\times10^{4}$ planets per episode.

\emph{Staying habitable.}  Life on Earth appeared within 0.2--0.5~Gyr of the planet's formation~\cite{2015PNAS..11214518B,2024NatEE...8.1654M}.  A planet must therefore stay habitable for an abiogenesis time $t_{\rm abi}=0.1$--$1$~Gyr, a factor of two around the terrestrial value, plus 1~Gyr over which life must persist, our definition of a sustained biosphere.  In a nuclear starburst of $1$--$10\,\Msun\,{\rm yr^{-1}}$ within 100~pc, thousands of supernovae per gigayear explode within the 8~pc at which the ozone layer is destroyed~\cite{2003ApJ...585.1169G}; gamma-ray bursts (GRBs) render most compact, strongly star-forming galaxies at $z\gtrsim3$ inhospitable~\cite{2014PhRvL.113w1102P,2015ApJ...810...41L}, with further GRBs from mergers of compact objects embedded in AGN disks~\cite{2026ApJ...998..171Z}; and a $10^{5}$--$10^{8}\,\Msun$ black hole that keeps growing as an AGN~\cite{2026JHEAp..5200576W} strips Earth-like atmospheres out to $\sim0.02$--$0.6$~kpc~\cite{2018MNRAS.479..171F,2017NatSR...716626B}.  LRDs themselves emit about ten times less soft X-ray flux than ordinary AGN of the same H$\alpha$ luminosity~\cite{2024ApJ...974L..26Y} and lack the extended X-ray-dominated regions of quasars~\cite{2026ApJ..1000L..11R}, so the black hole sterilizes planets mainly if it later shines as an ordinary AGN.  A surface biosphere survives only where star formation and accretion stop soon after the planet forms.  Combining these results, we adopt $f_{\rm sust}=10^{-3}$--$0.1$.  The upper end is the $\sim10\%$ of galaxies that remain safe from lethal GRBs at $z\sim3$~\cite{2015ApJ...810...41L}; the lower end lies two orders of magnitude below, because LRD hosts are far more compact and star-forming than those galaxies, and GRBs already make 95\% of the inner 4~kpc of the Milky Way inhospitable~\cite{2014PhRvL.113w1102P}.  It is the largest reduction in the chain and leaves 580 planets per episode.

\section{From feedstock to life}
\label{sec:life}

\emph{A sustained supply.}  Prebiotic synthesis needs concentrated feedstock.  HCN polymerizes toward purines only above $\sim10^{-2}$~M~\cite{1966Sci...153...72S}, and nucleobases form in warm little ponds only with repeated delivery over many wet--dry cycles~\cite{2017PNAS..11411327P,2022ApJ...932....9P}.  We express this as a supply of $\gtrsim10^{7}\,{\rm kg\,yr^{-1}}$ sustained for $\gtrsim10^{8}$~yr, $\Mcrit=10^{15}$--$10^{16}$~kg, and count a planet only if its inherited supply reaches it:
\begin{equation}
D=\frac{\epsilon_{\rm planet}\,\forg\,\Fbio\,M_\odot}{\Mcrit}\geq1 .
\label{eq:dose}
\end{equation}
Here $\epsilon_{\rm planet}=10^{-9}$--$10^{-7}$ is the fraction of the organics in one solar mass of natal gas that reaches a temperate rocky surface, calibrated on the $10^{15}$--$10^{17}$~kg of exogenous organics delivered to the early Earth~\cite{1990Sci...249..366C,1992Natur.355..125C}, largely by comets and meteorites that preserve material inherited from the protostellar phase~\cite{2019ESC.....3.1792R,2019MNRAS.490...50D}, and bounded by the destruction of most of the inherited carbonaceous organics in the Earth's building blocks~\cite{2021SciA....7.3632L}.  Each LRD episode delivers $5\times10^{44}$ of its molecules ($4\times10^{43}$--$7\times10^{45}$) to the surfaces of its temperate rocky planets.  Because LRD gas is metal-poor, the median dose is $D\approx0.04$ (Fig.~\ref{fig:travel}B shows its spread between draws), and the planet-to-planet spread in organic content lets a fraction $f_{\rm pass}$ of the planets exceed the threshold, with a median over draws of 4\% at every LRD epoch, or 3.3\% when the diluted gas of mergers (Section~\ref{sec:travel}) is included.  This leaves 14 planets per episode.  Planets could instead make their own feedstock by atmospheric photochemistry, impacts, and hydrothermal activity, which may match or exceed exogenous delivery~\cite{2005Sci...308.1014T,2022ApJ...932....9P}; we count such planets separately, because their chemistry is independent of LRDs.

\emph{The chemistry in between.}  Several reaction steps lie between feedstock and biomolecules.  The best-developed route, cyanosulfidic chemistry, builds the precursors of nucleotides, amino acids, and lipids from HCN using UV light, sulfur species, and metal ions~\cite{2009Natur.459..239P,2015NatCh...7..301P}; the products must then be phosphorylated, polymerized, and enclosed in vesicles.  These steps set three planetary prerequisites: enough UV light, exposed land for wet--dry cycles, and reactive phosphorus (Appendix~\ref{sec:param_appendix}).  We adopt $f_{\rm chem}=f_{\rm UV}f_{\rm land}f_{\rm P}$, which gives $f_{\rm chem}\approx0.21$ for FGK hosts and $1.2\times10^{-3}$ for M hosts; M-dwarf planets, a fifth of the candidates before this step, become 0.1\% of them after it.  About two planets per episode remain.

\emph{Life.}  No laboratory has yet produced a self-sustaining, evolving replicator~\cite{2018CSHPB..1034801J}, and we represent this final step by $\Pabi$.  Other attempts to quantify abiogenesis separate the environmental prerequisites from an intrinsic probability in the same way~\cite{2016PNAS..113.8127S,2012PNAS..109..395S}.  Each LRD episode thus forms $2.2\,\Pabi$ planets able to sustain life fed by LRD chemistry (16th--84th percentile $0.04$--$31$), with life appearing 0.1--1~Gyr after the planet forms.

\section{Living worlds across cosmic time}
\label{sec:epochs}

\begin{figure}
\centering
\includegraphics[width=\textwidth]{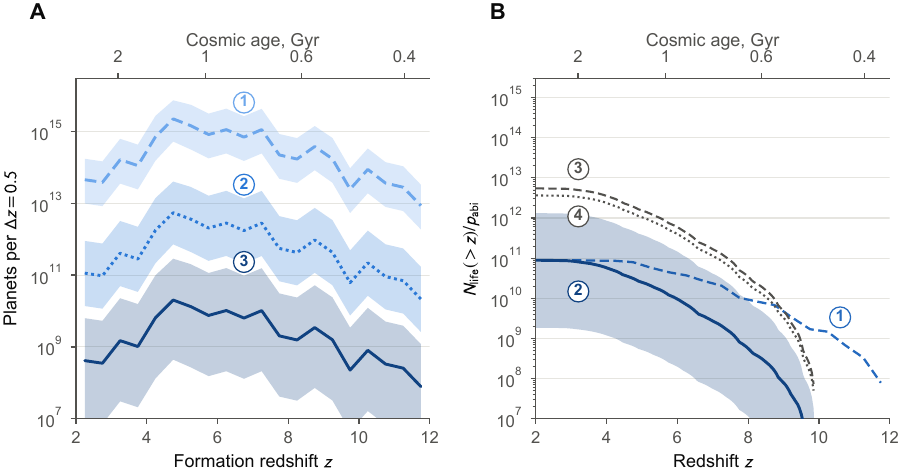}
\caption{Living worlds across cosmic time, in a comoving volume equal to the observable Universe.  Curves are medians and bands the 16th--84th percentiles.  (A) Planets per bin of formation redshift, $\Delta z=0.5$: 1, temperate rocky planets formed from LRD gas; 2, those that stay habitable for more than a gigayear; 3, those born able to sustain LRD-fed life, per unit $\Pabi$.  (B) Cumulative number per unit $\Pabi$ above redshift $z$: 1, planets born able to sustain LRD-fed life, by formation redshift (median); 2, those on which life has already emerged, a time $t_{\rm abi}=0.1$--$1$~Gyr (log-uniform) after formation; 3 and 4, the same for a subsurface biosphere and for endogenous chemistry alone.  The top axes give the cosmic age.}
\label{fig:epochs}
\end{figure}

In a comoving volume equal to the observable Universe, $V_{\rm obs}=1.13\times10^{4}\,\cGpc^3$~\cite{1999astro.ph..5116H}, the observed LRD abundance corresponds to $4.2\times10^{10}$ LRD episodes, each converting $\Delta t_{\rm bin}/t_{\rm LRD}$ generations of the observed reservoir into stars (Appendix~\ref{sec:param_appendix}).  They form $9.7\times10^{15}$ temperate rocky planets, of which $2.4\times10^{13}$ remain habitable, $5.9\times10^{11}$ receive a sustained inherited feedstock, and
\begin{equation}
N_{\rm life}=8.9\times10^{10}\,\Pabi
\label{eq:total}
\end{equation}
can sustain LRD-fed life (16th--84th percentile $1.8\times10^{9}$--$1.3\times10^{12}$; the 95th percentile is $5.8\times10^{12}$), of which $2\times10^{10}\,\Pabi$ form from gas mixed in mergers.  The earliest of these planets form with the earliest LRDs detected.  The bin at $z=11.5$--$12$, 0.37~Gyr after the Big Bang, already contributes $7.9\times10^{7}\,\Pabi$; $1.5\times10^{9}\,\Pabi$ form before $z=10$ (0.47~Gyr), $8.3\times10^{9}\,\Pabi$ (10\%) before $z=8$ (0.64~Gyr), and $3.7\times10^{10}\,\Pabi$ before $z=6$ (0.93~Gyr), while 77\% form at $z\approx4.5$--$7.5$ (Fig.~\ref{fig:epochs}A; Appendix~\ref{sec:results_appendix}).  Life emerges on these planets $t_{\rm abi}=0.1$--$1$~Gyr later (Section~\ref{sec:life}); taking $t_{\rm abi}$ to be distributed log-uniformly among planets, the first biospheres appear at $z\approx10$, 0.47~Gyr after the Big Bang, and life has emerged on $6\times10^{8}\,\Pabi$ planets by $z=8$, on $1.0\times10^{10}\,\Pabi$ (11\%) by $z=6$, and on 42\% of them by $z=4.5$ (Fig.~\ref{fig:epochs}B).  Because we take the metallicity of LRD gas to be the same at all redshifts (Section~\ref{sec:birth}) and nitrogen is available promptly, the fraction of planets that receive a sustained feedstock is the same at every epoch, and the earliest detected LRDs are as productive per unit gas as the later ones; this equality follows from the assumption.  The count depends more strongly on $Z_{\rm LRD}$ than on any other parameter (Section~\ref{sec:discussion}): draws with $Z_{\rm LRD}=0.03$--$0.045\,Z_\odot$ give 0.04 times the median number per episode, so if the gas of LRDs above $z\approx7$ lay at the low end of the adopted range, the $2.1\times10^{10}\,\Pabi$ planets formed before $z=7$ would fall to $\sim9\times10^{8}\,\Pabi$.  The large total reflects the size of the volume.  With the same planet-formation parameters, the cosmic star-formation history forms $\approx1.7\times10^{19}$ temperate rocky planets before $z=4.5$~\cite{2014ARAandA..52..415M}; LRD gas forms $5\times10^{-4}$ of them, and LRD-fed living worlds make up a fraction $5\times10^{-9}\,\Pabi$.

These counts include only the LRDs of the Kocevski et~al.~\cite{2025ApJ...986..126K} catalog, at $z=2.4$--$12$.  LRDs above $z\approx12$, if they exist, would add earlier worlds that current selections cannot reveal (Section~\ref{sec:birth}).  LRDs exist down to $z\approx0.1$~\cite{2026ApJ...997..364L,2026arXiv260521574L}, but their abundance below $z\approx2.4$ is not yet known.  Current estimates range from a lower limit of $7.5\times10^{-10}\,\cMpc^{-3}$ at $z=0.2$--$0.9$~\cite{2026arXiv260521574L} to values at $z\approx1.5$--$2.5$ that equal or exceed those at $z\approx5$~\cite{2026AandA...711A..24E,2026arXiv260630253L}, and current selections miss many LRDs~\cite{2026arXiv260407138R}.  The counts are therefore lower limits.  Two variants give higher counts.  If a subsurface biosphere suffices, so that only ocean loss sterilizes a planet~\cite{2018MNRAS.479..171F,2016AandA...592A..96G}, we take $f_{\rm sust}=0.3$--$1$, because quasar irradiation removes an Earth-like ocean from only 0.2\% of all planets and from a larger fraction in galaxy spheroids~\cite{2018MNRAS.479..171F}, and the count rises to $5.5\times10^{12}\,\Pabi$.  If endogenous synthesis alone can sustain prebiotic chemistry, $3.7\times10^{12}\,\Pabi$ planets formed from LRD gas qualify, although their chemistry is then independent of LRDs.

\section{The Milky Way today}
\label{sec:milkyway}

\begin{figure}
\centering
\includegraphics[width=\textwidth]{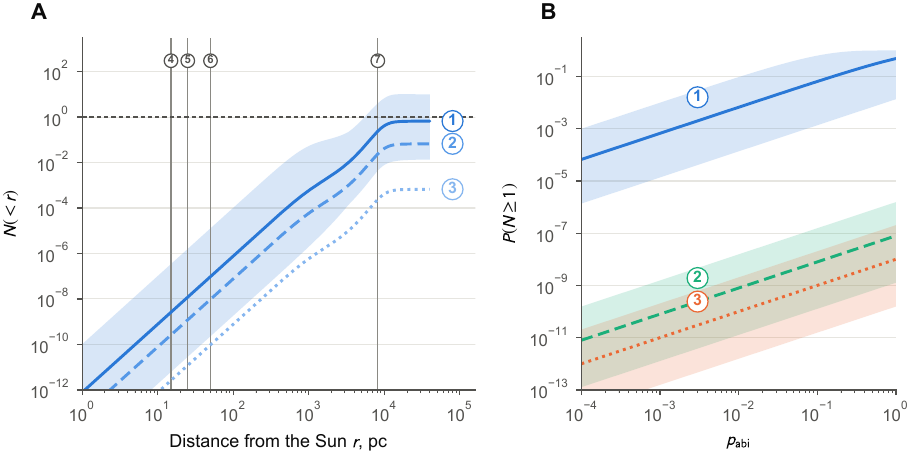}
\caption{The Milky Way today.  (A) Expected number $N(<r)$ of LRD-fed living worlds around stars still shining within a distance $r$ of the Sun: 1, $\Pabi=1$ (the band spans the 16th--84th percentiles and the two density profiles of the Galaxy's oldest stars); 2, $\Pabi=0.1$; 3, $\Pabi=10^{-3}$.  Vertical lines mark the reach of searches: 4, JWST spectra of rocky planets ($\lesssim15$~pc); 5, HWO targets (25~pc); 6, radio SETI limits (50~pc); 7, the Galactic Centre (8.2~kpc).  The dashed line marks one expected world.  (B) Probability of at least one LRD-fed living world as a function of $\Pabi$: 1, anywhere in the Milky Way; 2, within 50~pc; 3, within 25~pc of the Sun.}
\label{fig:milkyway}
\end{figure}

\emph{How many in a galaxy like ours.}  We apportion LRD episodes to present-day galaxies by stellar mass, taking a stellar mass density of $(2.3$--$3.5)\times10^{8}\,\Msun\,{\rm Mpc^{-3}}$ today~\cite{2012MNRAS.421..621B,2014ARAandA..52..415M} and a Milky Way of $(5$--$6)\times10^{10}\,\Msun$~\cite{2015ApJ...806...96L,2016ARAandA..54..529B}.  A Milky-Way-mass galaxy then descends from $0.7$ ($0.5$--$1.1$) LRD episodes at $z\geq2.4$ and contains $1.5\,\Pabi$ ($0.03$--$23$) planets born able to sustain LRD-fed life.  These planets are 11--13.4~Gyr old, most of them 12.4--13.1~Gyr.  At that age the main-sequence turnoff of metal-poor stars is $\approx0.8\,\Msun$, so for a Chabrier~\cite{2003PASP..115..763C} IMF 44\% of the FGK hosts, and all M-dwarf hosts, are still shining; since chemistry has removed nearly all M-dwarf planets, $0.67\,\Pabi$ ($0.014$--$10$) of these worlds orbit a living star today.  The probability that the Milky Way hosts at least one is 0.49 for $\Pabi=1$ (16th--84th percentile $0.014$--$1.00$), 0.065 for $\Pabi=0.1$, and $7\times10^{-4}$ for $\Pabi=10^{-3}$ (Fig.~\ref{fig:milkyway}B, Table~\ref{tab:mw}).

\emph{Where they are.}  Galaxy mergers move these planets together with their host stars and leave their number unchanged.  In simulated Milky-Way analogues more than half of the oldest stars end up outside the solar circle, and most of the stars formed before $z\sim5$ were accreted with their progenitors~\cite{2017MNRAS.465.2212S,2018MNRAS.480..652E}, while dense cores formed at $z\gtrsim7$ survive as the central $\sim100$~pc of present-day galaxies~\cite{2023ApJ...955L..12B}, and dense satellites survive accretion better than diffuse ones~\cite{2007MNRAS.374.1227B}.  In the Milky Way itself, the oldest metal-poor stars form a centrally concentrated population, described by a Gaussian of $\sigma\approx2.7$~kpc~\cite{2022ApJ...941...45R} or a Plummer profile of scale 3.5~kpc~\cite{2025MNRAS.537.3730H}, and make up only 0.09--0.29\% of stars near the solar radius~\cite{2018MNRAS.480..652E}.  Integrating these profiles over spheres around the Sun (Fig.~\ref{fig:milkyway}A), the 25~pc volume holds a fraction $2\times10^{-9}$--$1.2\times10^{-7}$ of the population.  For $\Pabi=1$, the expected number stays far below one within any distance that current searches reach and is 0.7 for the Galaxy as a whole.

\emph{Comparison with searches.}  The expected number within 25~pc of the Sun is $1.0\times10^{-8}\,\Pabi$.  This volume contains the 164 target stars of the Habitable Worlds Observatory (HWO)~\cite{2024ApJS..272...30H}, and about 3600 stars and brown dwarfs lie within 20~pc~\cite{2024ApJS..271...55K}.  Within 50~pc, the reach of the deepest radio limits of the search for extraterrestrial intelligence (SETI)~\cite{2020MNRAS.498.5720W}, the expected number is $8\times10^{-8}\,\Pabi$.  No temperate rocky planet has yet been characterized well enough to test for biosignatures~\cite{2024ApJ...970L...2C}; the one claimed biosignature, on the sub-Neptune K2-18\,b, is not confirmed~\cite{2025ApJ...983L..40M,2025AandA...700A.284L}; and radio searches have covered only $\sim6\times10^{-18}$ of their parameter space~\cite{2018AJ....156..260W}.  Given the Earth, the probability that no LRD-fed biosphere lies within 25~pc is $1-1.0\times10^{-8}\,\Pabi$, and within the 50~pc reach of radio SETI it is $1-8\times10^{-8}\,\Pabi$.  The observed situation, one inhabited planet in the Milky Way and none detected near the Sun, is thus the expected outcome.  The Galaxy may still host a biosphere fed by LRD chemistry, with probability 0.49 for $\Pabi=1$, but it would orbit one of the Galaxy's oldest stars, in the inner kiloparsecs or the stellar halo.  These regions lie beyond the reach of all biosignature searches performed or planned; only technosignature searches toward the Galactic Centre~\cite{2021AJ....162...33G} probe the inner Galaxy.

\emph{An LRD phase in the Galaxy's past.}  Our own Galaxy may have passed through the phase we have followed, either because its main progenitor was an LRD or because it absorbed a galaxy that was one.  Ruffini and Wang~\cite{2026ApJ..1000L..11R} noted that an LRD is structurally equivalent to the innermost bulge of the Milky Way, where the CMZ and the nuclear star cluster reside, and low-spin dark-matter halos produce such compact cores~\cite{2025ApJ...989L..19P}.  Several observations point in this direction.  The mass of Sgr~A$^*$, $4.1\times10^{6}\,\Msun$~\cite{2008ApJ...689.1044G}, lies in the range of LRD black holes, and a seed of $\sim10^{6}\,\Msun$ formed at $z\sim20$--$30$~\cite{2026JHEAp..5200576W} would have grown only by a factor of a few since the LRD epoch.  Before the Galactic disk formed, at $[{\rm Fe/H}]\lesssim-1.5$, up to 50--70\% of the Galaxy's in-situ stars formed in massive bound clusters with nitrogen-to-oxygen ratios as high as in the $z\approx11$ galaxy GN-z11~\cite{2023MNRAS.525.4456B}, and this ancient population is compact and centrally concentrated~\cite{2022ApJ...941...45R}.  Its metallicity and compactness match those of LRD hosts (Section~\ref{sec:birth}).  Progenitors of Milky-Way analogues observed with JWST already contain $10^{9.3}\,\Msun$ of stars within 2~kpc at $2<z<5$~\cite{2025ApJ...994...94T}, so an LRD phase of the main progenitor, with a host of $10^{7.5}$--$10^{8.5}\,\Msun$, would have occurred at $z\gtrsim6$.  The galaxies the Milky Way absorbed match LRD hosts at lower redshift.  Kraken, Gaia-Enceladus, Sequoia, the Helmi streams, and Sagittarius had $(0.6$--$4.6)\times10^{8}\,\Msun$ of stars when they were accreted~\cite{2018Natur.563...85H,2018MNRAS.478..611B,2020MNRAS.498.2472K}, and a structure of $\sim5\times10^{8}\,\Msun$ lies buried within 4~kpc of the Galactic Centre~\cite{2021MNRAS.500.1385H}.  Our apportionment gives a Milky-Way-mass galaxy $0.7$ ($0.5$--$1.1$) LRD episodes, so the probability that the Galaxy or one of its progenitors was an LRD is $1-e^{-0.7}\approx0.5$ ($0.4$--$0.7$).  This is a lower limit, because it counts only LRDs at $z\geq2.4$; galaxies accreted late, such as Sagittarius, could have passed through an LRD phase at lower redshift.  A large fraction of the stars formed before $z\sim5$ formed in progenitors that were later accreted~\cite{2017MNRAS.465.2212S,2018MNRAS.480..652E}, so an in-situ LRD phase and an accreted LRD are comparably likely.  In the first case the Galaxy's LRD-fed worlds orbit the nitrogen-rich stars of its ancient core; in the second they orbit accreted stars in the inner Galaxy and the stellar halo.

\emph{What the Earth inherited.}  LRDs are not confined to the early Universe.  Analogues are found down to $z\approx0.1$~\cite{2025ApJ...980L..34L,2026ApJ...997..364L,2026MNRAS.545f2235J,2026arXiv260514233P,2026AandA...709L..11D}, including an LRD-like galaxy at $z=0.433$~\cite{2026ApJ...999...30C}, the epoch at which the Sun formed.  The Sun formed 4.57~Gyr ago~\cite{2012Sci...338..651C}, at $z\approx0.42$, in the Galactic disk at 5--7~kpc from the Galactic Centre~\cite{1996AandA...314..438W,2013AandA...558A...9M,2023MNRAS.526.6088B}, within a cluster of $10^{3}$--$10^{4}$ stars~\cite{2010ARAandA..48...47A}, and has since migrated outward to its present distance of 8.2~kpc~\cite{2019AandA...625L..10G}.  Our count includes only planets formed in the gas of LRDs at $z\geq2.4$, at least 6~Gyr before the Sun, because the abundance of LRDs at lower redshift is not yet known (Section~\ref{sec:epochs}).  This is a limitation of the calculation, not a statement that the Earth is unrelated to LRDs: LRDs at lower redshift were active when the Sun formed, and the Earth may be connected to them, as well as to the Galaxy's own LRD past.  We estimate four such connections in Appendix~\ref{sec:earth_appendix}.  The Sun could have formed in the gas of an LRD active at its birth: the abundance of low-redshift LRDs gives a probability between $10^{-8}$ and $5\times10^{-4}$ for a star born at that epoch, lower for a star born in a galactic disk.  If the Galaxy was an LRD or absorbed one, a fraction $10^{-4}$--$4\times10^{-3}$ of the Earth's carbon, nitrogen, and oxygen atoms passed through an LRD host, and refractory organics carried on dust grains delivered $10^{24}$--$10^{33}$ LRD-made molecules, $\lesssim10^{-7}$ of the exogenous organics the Earth received.  Finally, Ruffini and Wang~\cite{2026ApJ..1000L..11R} proposed that the Galactic Centre, whose clouds have a chemistry similar to that of comets, contributed to the building blocks of life delivered to the early Earth.  The Sun was born closer to the Galactic Centre than it is now, and the CMZ, the Galaxy's present LRD-like core, exports its chemistry in the Galactic nuclear wind; it may have supplied $10^{-5}$--$3\times10^{-2}$ of the interstellar organics that the Solar System inherited.  This component would carry the Galactic Centre's isotopic signature, $^{12}$C/$^{13}$C$\,\approx20$--$25$ and a low deuterium content~\cite{1994ARAandA..32..191W,2000Natur.405.1025L}, which analyses of the organics returned from Ryugu and Bennu can test.

\begin{table}
\renewcommand{\baselinestretch}{1}\selectfont
\centering
\caption{The Milky Way and the searches}
\label{tab:mw}
\footnotesize
\setlength{\tabcolsep}{8pt}\renewcommand{\arraystretch}{1.3}
\begin{tabular}{lIcIc}
\toprule
Quantity & $\Pabi=1$ & $\Pabi=0.1$ \\
\midrule
LRD episodes in the Galaxy's past & \multicolumn{2}{c}{0.7 (0.5--1.1)} \\
LRD-fed living worlds ever born & 1.5 & 0.15 \\
\quad around stars still shining & 0.67 & 0.067 \\
$P(\geq1$ in the Galaxy today$)$ & 0.49 & 0.065 \\
Expected within 50~pc (radio SETI) & $8\times10^{-8}$ & $8\times10^{-9}$ \\
Expected within 25~pc (HWO targets) & $1.0\times10^{-8}$ & $1.0\times10^{-9}$ \\
Expected within 15~pc (JWST rocky planets) & $2.2\times10^{-9}$ & $2.2\times10^{-10}$ \\
\bottomrule
\end{tabular}

\medskip
\begin{flushleft}\footnotesize Medians, counting LRDs at $z\geq2.4$ only; for $\Pabi=10^{-3}$ the probability of at least one in the Galaxy today is $7\times10^{-4}$.  No biosignature or technosignature has been detected in any of these volumes.\end{flushleft}
\end{table}

\section{What controls the answer, and how to test it}
\label{sec:discussion}

\emph{The weakest links.}  We rank the sampled parameters by their rank correlation with the total count (Appendix~\ref{sec:results_appendix}).  The metallicity of LRD gas dominates (rank correlation $+0.57$), because it sets both the organic fraction and the formation of rocky planets; it is followed by the sustained-supply threshold, $\Mcrit$ ($-0.45$), the survival of habitability in the compact, active environment, $f_{\rm sust}$ ($+0.43$), the organic fraction ($+0.22$), the metallicity threshold for planet formation ($-0.21$), the star-formation efficiency of the reservoir and its cold gas mass ($+0.19$ each), and disk survival ($+0.15$).  The outflow and merger parameters have correlations of $|r|\leq0.01$: gas returned by outflows stays below the sustained-supply threshold, and merger mixing, which dilutes the organics while adding planets, changes the total by 7\%.  The count is therefore limited by two properties of LRDs.  Their gas is metal-poor, which limits both its organics and its planets, and their compact, active nuclei make long-lived habitability rare.  A subsurface biosphere would raise the count by a factor of $\sim60$, and the count scales linearly with $\Pabi$.

\emph{The role of LRDs.}  Ruffini and Wang~\cite{2026ApJ..1000L..11R} proposed that LRDs, which may make up a few tens of percent of early galaxies, seeded later generations of galaxies with prebiotic material.  Our results quantify each route of this picture.  Planets formed from LRD gas during the episode receive a sustained feedstock in a few percent of cases; gas mixed in mergers forms about a quarter of the LRD-fed worlds; gas thrown into tidal features during mergers keeps 1--50\% of its ice-locked organics; gas returned by outflows carries $\sim10^{-8}$ of the sustained supply a planet needs; and refractory organics on dust grains reach other galaxies and later planets, including the Earth.  With $4\times10^{10}$ LRD episodes in the observable-Universe volume, each route operates many times, so rare outcomes along any of them add up.  And if endogenous synthesis is as productive on young rocky planets as models of the early Earth suggest~\cite{2005Sci...308.1014T,2022ApJ...932....9P}, inherited organics do not limit the emergence of life, and the main role of LRDs is to host an early generation of planets.  Because the elements of life were in place by $z\approx12$, when the earliest LRDs detected already existed, planets able to sustain LRD-fed life formed within the first 0.4~Gyr of cosmic history, and life could emerge on them from 0.5~Gyr; earlier LRDs, beyond the reach of current selections, would move both epochs earlier.

\emph{Tests.}  Each step can be tested.  Metallicities of LRDs above $z\approx7$ would test whether the earliest LRDs are as productive as the later ones, and the machine-learning methods developed for quasar spectra and absorption lines~\cite{2022MNRAS.511.4490R,2024MNRAS.533.1976M,2025ApJS..276...37L} can be adapted to the growing samples of LRD spectra.  Deeper ALMA CO, [C\,{\sc ii}], and dust observations can confirm or rule out the cold reservoirs of LRD hosts~\cite{2025ApJ...990L..61C,2026ApJ...997..218A}, and searches for complex organics in low-metallicity gas constrain $\forg$.  Clustering and duty-cycle measurements constrain $t_{\rm LRD}$, and outflow signatures the fraction of the reservoir that forms stars.  The occurrence of rocky planets around the Galaxy's oldest, most metal-poor stars tests the planet-formation step directly, and continued atmospheric characterization of rocky planets tests the habitability and chemistry steps.  Local biosignature searches cannot test the model directly, because it predicts no detection within their reach.  Three observations bear on the Milky Way's LRD past.  The Roman microlensing survey of the bulge will measure the occurrence of planets around the nitrogen-rich and accreted stars of the inner Galaxy~\cite{2019ApJS..241....3P}, technosignature searches toward the Galactic Centre look where LRD-fed worlds would be concentrated~\cite{2021AJ....162...33G}, and isotopic analyses of the organics returned from Ryugu and Bennu can reveal a Galactic-Centre component.

\section{Conclusions}
\label{sec:conclusions}

We have followed the prebiotic molecules of JWST Little Red Dots from their synthesis to the planets on which they might have fed life, giving each step the probability supported by published work.

\emph{Birth.}  LRDs are detected from $z\approx12$, 0.37~Gyr after the Big Bang, and in the JWST sample are most common at $z\approx4.5$--$7.5$, $(4$--$9)\times10^{-5}\,\cMpc^{-3}$; their abundance at lower redshift is not yet known.  Each LRD episode assembles $\sim10^{55}$ prebiotic molecules in cold host gas of 3--20\% of the solar metallicity, and the CHNOPS elements are already in place at the earliest epochs.

\emph{Travel.}  These molecules are short-lived outside dense ices, and they travel by four routes.  Planets that form from LRD gas during the 10--30~Myr LRD phase receive them from comets and meteorites; in a third of episodes a merger mixes LRD gas with a companion's while its organics survive, and about a quarter of the LRD-fed worlds form from this mixed gas, while mergers can also throw LRD gas into tidal features, which 1--50\% of its ice-locked organics reach intact; outflows return LRD gas after 0.1--1~Gyr with $\sim10^{-8}$ of the sustained supply a planet needs; and refractory organics on dust grains reach later planets, including the Earth.

\emph{Planets and life.}  Of the $\sim3\times10^{5}$ temperate rocky planets formed per episode, $\sim600$ survive their compact nuclear birthplace and stay habitable for a gigayear, $\sim14$ receive a sustained inherited feedstock, and $\sim2$ also have the UV light, exposed land, and phosphorus that prebiotic chemistry needs.  Each of them then forms long-lasting life with a probability $\Pabi=10^{-3}$--$1$.

\emph{Across cosmic time.}  A comoving volume equal to the observable Universe contains $9\times10^{10}\,\Pabi$ LRD-fed living worlds (16th--84th percentile $1.8\times10^{9}$--$1.3\times10^{12}$), a lower limit because LRDs below $z\approx2.4$ and above $z\approx12$ are not included.  The first of these planets form with the earliest LRDs detected, at $z\approx12$, $1.5\times10^{9}\,\Pabi$ before $z=10$, and 77\% at $z\approx4.5$--$7.5$, 0.7--1.3~Gyr after the Big Bang.  Life emerges on them 0.1--1~Gyr later, from $z\approx10$ onward, and by $z=6$ on 11\% of them.  The number formed at the earliest epochs rests on the assumption that the metallicity of LRD gas does not change with redshift, which has not been tested above $z\approx7$.

\emph{Today.}  A galaxy like the Milky Way should host $\sim0.7\,\Pabi$ of them around stars still shining; for $\Pabi=1$ the probability that at least one exists today is 0.49.  Only $1.0\times10^{-8}\,\Pabi$ lie within 25~pc, so a Milky Way that contains the Earth and no detected second inhabited planet near the Sun is the expected outcome.  The Milky Way may itself have been an LRD, or may have absorbed one; we estimate the probability at no less than one half.  In that case the Earth probably contains atoms processed in LRD hosts and a trace of LRD-made molecules.

\begin{acknowledgments}
We thank the JWST teams whose public catalogs make this type of calculation possible.
Anthropic's Claude was used to assist with language editing and code development. All AI-assisted text and code were reviewed by the authors, who take full responsibility for the content of this paper.
\end{acknowledgments}

% \section*{Funding}
% (to be completed by the authors)

\section*{Conflict of interest}

The authors declare that they have no conflicts of interest.

\appendix

\section{LRD density: details}
\label{sec:density_appendix}

\begin{figure}
\centering
\includegraphics[width=\textwidth]{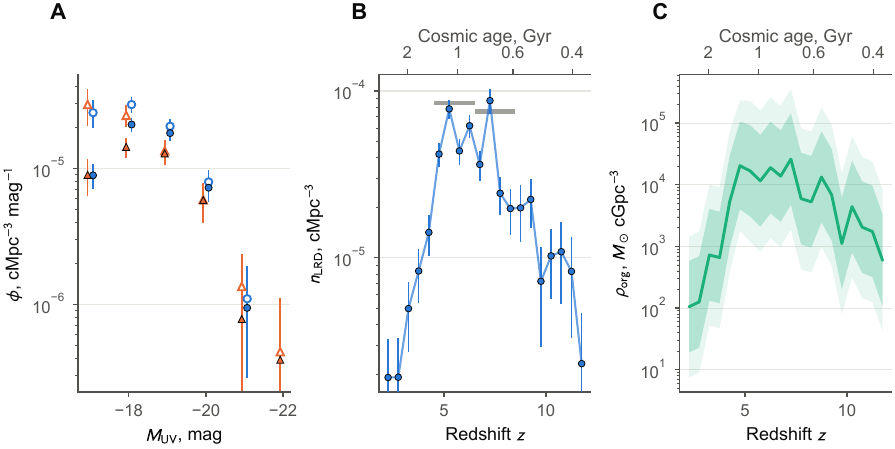}
\caption{Birth.  (A) UV luminosity function of the 341 LRDs used here, from our coherent $1/V_a$ weights (filled symbols), in two redshift bins: circles, $4.5<z<6.5$; triangles, $6.5<z<8.5$.  Open symbols are the values published by Kocevski et~al.~\cite{2025ApJ...986..126K}.  (B) Comoving number density of LRDs in bins of $\Delta z=0.5$ with Poisson errors; gray bars are the integrated luminosity function of Kocevski et~al.~\cite{2025ApJ...986..126K}.  (C) Mass density of prebiotic organics in LRD host gas.  In (C) the solid curve is the median and the dark and light bands the 16th--84th and 5th--95th percentiles.}
\label{fig:birth}
\end{figure}

For each source we use the redshift $z_i$, field $f(i)$, F444W magnitude $m_i$, rest-optical slope $\beta_{{\rm opt},i}$, and UV absolute magnitude $M_{{\rm UV},i}$ from Kocevski et~al.~\cite{2025ApJ...986..126K}.  With $f_\nu\propto\nu^{\alpha_{\nu,i}}$ and $\alpha_{\nu,i}=-(\beta_{{\rm opt},i}+2)$, the F444W magnitude of source $i$ moved to redshift $z$ is
\begin{equation}
\begin{split}
m_i(z)=&\,m_i+5\log_{10}\left[\frac{D_L(z)}{D_L(z_i)}\right]\\
&-2.5(1+\alpha_{\nu,i})\log_{10}\left[\frac{1+z}{1+z_i}\right],
\end{split}
\label{eq:magmove}
\end{equation}
the luminosity-distance scaling plus the power-law $K$-correction~\cite{1999astro.ph..5116H,2002astro.ph.10394H}.  The selection limit of field $f$ is the magnitude at which $\mathrm{S/N}=12$, $m_{{\rm lim},f}=m_{5\sigma,f}-2.5\log_{10}(12/5)$, and $z_{{\rm max},i,f}$ solves $m_i(z)=m_{{\rm lim},f}$.  Table~\ref{tab:fields} lists the field areas and limits.  Figure~\ref{fig:birth} compares the resulting UV luminosity function with the published one and shows the number density of LRDs and the mass density of their prebiotic organics as functions of redshift.

\begin{table}
\renewcommand{\baselinestretch}{1}\selectfont
\centering
\caption{Field inputs for the $1/V_a$ weights~\cite{2025ApJ...986..126K}}
\label{tab:fields}
\footnotesize\setlength{\tabcolsep}{8pt}\renewcommand{\arraystretch}{1.3}
\begin{tabular}{lIcIcIcIc}
\toprule
Field & Area, arcmin$^2$ & $m_{5\sigma}$, mag & $m_{\rm lim}$, mag & LRDs\\
\midrule
NGDEEP & 11.4 & 30.65 & 29.70 & 10 \\
UNCOVER & 43.8 & 29.33 & 28.38 & 23 \\
JADES & 62.1 & 29.62 & 28.67 & 46 \\
CEERS & 88.6 & 28.97 & 28.02 & 64 \\
PRIMER-COS & 138.9 & 28.78 & 27.83 & 81 \\
PRIMER-UDS & 243.0 & 28.54 & 27.59 & 117 \\
\bottomrule
\end{tabular}

\medskip
\begin{flushleft}\footnotesize $m_{5\sigma}$ is the F444W $5\sigma$ point-source depth; $m_{\rm lim}$ is the $\mathrm{S/N}=12$ selection limit (AB magnitudes).\end{flushleft}
\end{table}

\section{CHNOPS availability}
\label{sec:gates_appendix}

Carbon, oxygen, sulfur, and phosphorus are produced promptly by massive stars; nitrogen, whose detection at $z=12$--$14$ requires enrichment within tens of Myr~\cite{2023MNRAS.523.3516C,2024ApJ...972..143C,2026OJAp....956033N,2026MNRAS.tmp.1597R}, is given a timescale $\tau_{\rm N}=10$--$40$~Myr.  The CHNOPS factor is $\Fbio=1-\exp(-t_{\rm enr}/\tau_{\rm N})$, with $t_{\rm enr}$ the time since the first star formation at $z=20$; it exceeds 0.99 at $z<12$.  The overall metallicity enters the model through $\forg$ and $F_Z$.

\section{Derivation of the adopted ranges}
\label{sec:param_appendix}

\emph{Counting episodes and mergers.}  Each LRD consumes one reservoir per episode, so a redshift bin of duration $\Delta t_{\rm bin}$ turns $\Delta t_{\rm bin}/t_{\rm LRD}$ generations of the observed standing reservoir into stars, and the number of temperate rocky planets formed per bin is $N_{\rm TTP}=\eta_{\rm TTP}F_Z\epsilon_\star\rho_{\rm mol}V_{\rm obs}\Delta t_{\rm bin}/t_{\rm LRD}$, with $\rho_{\rm mol}$ the cold-gas density of LRD hosts and $\eta_{\rm TTP}=\eta_{\rm FGK}+f_M\eta_M$.  A merger mixes LRD gas with its partner's before the LRD organics are reset only if it occurs within $t_{\rm LRD}+\tau_{\rm ice}$, which happens with probability $P_{\rm mix}=1-\exp[-R_{\rm merge}(t_{\rm LRD}+\tau_{\rm ice})]\approx0.36$ ($0.26$--$0.48$).  In a merger of mass ratio 1:3 to 1:10 the partner contributes 10--25\% of the combined mass, and gas-rich mergers funnel part of its gas into the nucleus~\cite{1996ApJ...471..115B,2009ApJ...691.1168H}; we therefore take the LRD share of the mixed gas as $\mu_{\rm mix}=0.6$--$0.8$.  Mixing lowers the organic fraction by $\mu_{\rm mix}$ but raises the number of planets forming from LRD-bearing gas by $1/\mu_{\rm mix}$: merger mixing multiplies the number of planets by $(1-P_{\rm mix})+P_{\rm mix}/\mu_{\rm mix}$ and evaluates $f_{\rm pass}$ at dose $\mu_{\rm mix}D$ for the mixed fraction.

\emph{Organic fraction.}  The mass fraction is $\forg=\sum_iX_im_i/(2.8\,m_{\rm u})$, with $X_i$ the abundance relative to H$_2$ and $2.8\,m_{\rm u}$ the mass per H$_2$ including helium.  For the prebiotic subset in G+0.693$-$0.027 this gives $4.7\times10^{-7}$ (strictest subset) to $1.2\times10^{-6}$~\cite{2008ApJ...672..352R,2018MNRAS.478.2962Z,2020AsBio..20.1048J}, which we adopt, rounded, as $f_{{\rm org},\odot}=5\times10^{-7}$--$1.2\times10^{-6}$.  It is scaled linearly with $Z_{\rm LRD}$ and multiplied by the core-to-core depletion $\delta$, which we treat as a planet-to-planet distribution in $f_{\rm pass}$.

\emph{Molecular clocks.}  The unshielded photodissociation rates of CH$_3$OH, CH$_3$CHO, and NH$_2$CHO in the interstellar radiation field are $(1.4$--$2.7)\times10^{-9}\,{\rm s^{-1}}$, and their cosmic-ray-induced destruction rates in shielded gas are $(1.8$--$3.3)\times10^{-13}\,{\rm s^{-1}}$ at $\zeta=10^{-16}\,{\rm s^{-1}}$~\cite{2017AandA...602A.105H}.  Ice-locked glycine has a half-life of $(2.2$--$4.7)\times10^{7}$~yr in dense clouds~\cite{2001ApJ...550L..95E}.

\emph{Recycling dose.}  An $\MUV=-20$ LRD forms stars at $\dot M_\star\approx3\,\Msun\,{\rm yr^{-1}}$, from its UV luminosity for a Chabrier IMF~\cite{2014ARAandA..52..415M}, and $\dot M_\star$ scales with UV luminosity as $\Mmol$ does.  Recycled organics are diluted into the gas that forms stars over $t_{\rm rec}$, so their organic fraction is $\forg\Fbio\,\epsilon_{\rm launch}f_{\rm surv}f_{\rm rec}\,\Mmol/(\dot M_\star t_{\rm rec})$, with $\Mmol/(\dot M_\star t_{\rm rec})=0.01$--$1$.  A recycled dose $D=1$ would need a transport efficiency $\epsilon_{\rm launch}f_{\rm surv}f_{\rm rec}\sim270$ (16th--84th percentile 26--2900), far larger than unity.

\emph{Delivery and threshold.}  $\epsilon_{\rm planet}$ is the exogenous organic mass delivered to the early Earth, $10^{15}$--$10^{17}$~kg, divided by the prebiotic organics in one solar mass of gas with $\forg\approx5\times10^{-7}$ ($10^{24}$~kg).  $\Mcrit=10^{15}$--$10^{16}$~kg corresponds to $\gtrsim10^{7}\,{\rm kg\,yr^{-1}}$ for $\gtrsim10^{8}$--$10^{9}$~yr; filling the whole ocean to the $10^{-2}$~M HCN threshold would require $\approx4\times10^{17}$~kg, so local concentration in ponds is essential~\cite{2017PNAS..11411327P,2022ApJ...932....9P}.  The spreads of $\epsilon_{\rm planet}$ and $\delta$ are treated as planet-to-planet variation, and $f_{\rm pass}$ is the fraction of planets with $D\geq1$.

\emph{Planets.}  For a Chabrier~\cite{2003PASP..115..763C} IMF one solar mass of stars contains 0.19--0.22 FGK stars ($0.6$--$1.4\,\Msun$) and 1.2--1.4 M dwarfs; multiplying by the habitable-zone occurrence of Bryson et~al.~\cite{2021AJ....161...36B}, Dressing and Charbonneau~\cite{2015ApJ...807...45D}, and Hsu et~al.~\cite{2020MNRAS.498.2249H} gives $\eta_{\rm FGK}=(0.19$--$0.22)\times(0.37$--$0.60)=0.07$--$0.13$ and $\eta_M=(1.2$--$1.4)\times(0.16$--$0.33)=0.2$--$0.45$ per solar mass; we apply the rate that Bryson et~al.~\cite{2021AJ....161...36B} derive for stars of 4800--6300~K to all FGK stars.  For the integrated star-formation efficiency $\epsilon_{\rm int}=(1/0.77+\Sigma_{\rm crit}/\Sigma)^{-1}$ with $\Sigma_{\rm crit}=2800\,\Msun\,{\rm pc^{-2}}$~\cite{2018MNRAS.475.3511G}, $10^{7}$--$10^{8}\,\Msun$ within 100~pc ($\Sigma\approx300$--$3000\,\Msun\,{\rm pc^{-2}}$) gives 0.10--0.46; the CMZ forms stars an order of magnitude below density-based expectations, $\epsilon\sim0.03$--$0.1$ over 15--30~Myr~\cite{2013MNRAS.429..987L}.

\emph{Habitability.}  For one core-collapse supernova per $\sim100\,\Msun$ formed, $1$--$10\,\Msun\,{\rm yr^{-1}}$ within a 100~pc sphere gives $\sim2\times10^{-9}$--$2\times10^{-8}\,{\rm pc^{-3}\,yr^{-1}}$ and $5\times10^{3}$--$5\times10^{4}$ supernovae per gigayear within 8~pc.  The atmosphere-loss distance of Forbes and Loeb~\cite{2018MNRAS.479..171F} scales as $1.8\,{\rm kpc}\,(M_{\rm BH}/10^{9}\,\Msun)^{1/2}$, i.e.\ $\approx0.02$--$0.6$~kpc for $10^{5}$--$10^{8}\,\Msun$.  $f_{\rm sust}$ is therefore the fraction of planets whose surroundings quench before life must persist, bounded above by the $\sim10\%$ of galaxies hospitable at $z\sim3$~\cite{2015ApJ...810...41L}.

\emph{Chemistry.}  The best-developed route, cyanosulfidic chemistry, builds sugars and the precursors of nucleotides, amino acids, and lipids from HCN using UV light at 200--280~nm, sulfur species, and metal ions~\cite{2009Natur.459..239P,2015NatCh...7..301P}; a UV-free route to both purine and pyrimidine nucleotides runs on wet--dry cycles alone~\cite{2019Sci...366...76B}.  Nucleotides must then be phosphorylated and polymerized (on clays, oligomers of up to 55 units form with repeated feeding~\cite{1996Natur.381...59F}) and enclosed in fatty-acid vesicles, which favour fresh-water pools~\cite{2008Natur.454..122M,2020AsBio..20..429D}.  These steps set three planetary prerequisites.  The first is UV light.  Stars hotter than $\sim4400$~K provide enough quiescent UV, while cooler stars need frequent flares~\cite{2018SciA....4.3302R}, which only a minority of M dwarfs produce~\cite{2020AJ....159...60G}.  M-dwarf planets receive 100--1000 times less bioactive UV~\cite{2017ApJ...843..110R}, but the photochemistry is fast enough that the flare criterion is the stronger constraint~\cite{2021AsBio..21.1099R}.  The second is exposed land.  Wet--dry cycles need subaerial land, which survives only on planets with water contents below a few tenths of a percent by weight~\cite{2014ApJ...781...27C}; such water contents are 10--100 times rarer around M dwarfs than around G dwarfs~\cite{2015NatGe...8..177T,2022NatAs...6.1296K}, most habitable planets may be ocean worlds~\cite{2017MNRAS.468.2803S}, and the Hadean Earth itself may have had only volcanic islands~\cite{2018Life....8...55B,2021Life...11.1142K}.  The third is phosphorus.  Reactive phosphorus can come from meteoritic phosphides~\cite{2005AsBio...5..515P} and from carbonate-rich lakes that concentrate phosphate~\cite{2020PNAS..117..883T}, and stellar phosphorus tracks iron down to low metallicity~\cite{2014ApJ...797...69R,2014ApJ...796L..24J}, although planets of stars poor in phosphorus may have little of it at their surfaces~\cite{2020ApJ...900L..38H}.

\emph{Chemistry factors.}  Of the FGK hosts of 0.6--1.4~$\Msun$, 75\% are hotter than the 4400~K needed for enough quiescent UV~\cite{2018SciA....4.3302R}, and active K dwarfs can supply the rest, so $f_{\rm UV}=0.7$--$1$; among M dwarfs only the minority of fast rotators flare often enough~\cite{2020AJ....159...60G}, so $f_{\rm UV}=0.01$--$0.5$.  Around early-to-mid M dwarfs, 5--10\% of small planets have water contents suitable for exposed land~\cite{2022NatAs...6.1296K}, and Earth-like water contents are 10--100 times rarer there than around G dwarfs~\cite{2015NatGe...8..177T}; we take $f_{\rm land}=0.01$--$0.1$ for M hosts and $0.2$--$0.6$ for FGK hosts, below unity because large planets tend to become ocean worlds~\cite{2014ApJ...781...27C,2017MNRAS.468.2803S}.  Reactive phosphorus has two independent sources on land, meteoritic phosphides and carbonate-rich lakes~\cite{2005AsBio...5..515P,2020PNAS..117..883T}, so we take $f_{\rm P}=0.5$--$1$ for FGK hosts; around M dwarfs the UV-driven route from phosphides to phosphate~\cite{2020NatGe..13..344R} is weaker, and we take $f_{\rm P}=0.3$--$1$.

\emph{The Milky Way.}  The Milky Way's share of the stellar mass in $V_{\rm obs}$ is $M_{\rm MW}/(\rho_\star V_{\rm obs})$.  The main-sequence fraction uses the Chabrier~\cite{2003PASP..115..763C} number of stars between 0.6 and 0.8~$\Msun$ relative to 0.6--1.4~$\Msun$ (0.44).  The fraction of the ancient population within distance $r$ of the Sun is the integral of the Gaussian or Plummer profile over that sphere; the upper envelope near the Sun uses the counting estimate of $\approx2900$ stars within 20~pc, the $\approx3600$ objects of the census at a star-to-brown-dwarf ratio of 4:1~\cite{2024ApJS..271...55K}, of which 0.09--0.29\% formed before $z=5$~\cite{2018MNRAS.480..652E}, against $10^{8}$--$10^{9}\,\Msun$ of such stars in the Galaxy~\cite{2022ApJ...941...45R,2025MNRAS.537.3730H}, of which 1.4 stars per solar mass formed are still shining for a Chabrier~\cite{2003PASP..115..763C} IMF.

\section{Redshift-resolved results and sensitivity}
\label{sec:results_appendix}

\begin{figure}
\centering
\includegraphics[width=0.7\textwidth]{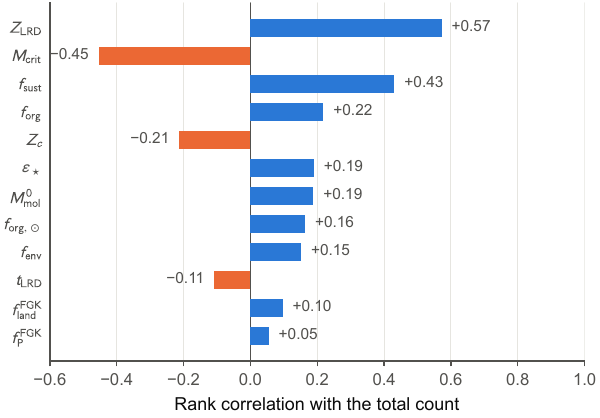}
\caption{What controls the count.  Rank correlation of each sampled parameter with the total number of LRD-fed living worlds; the twelve most influential parameters are shown.}
\label{fig:sensitivity}
\end{figure}

Table~\ref{tab:results} gives the journey by formation redshift, and Figure~\ref{fig:sensitivity} ranks the sampled parameters by their rank correlation with the total count of LRD-fed living worlds.

\begin{table}
\renewcommand{\baselinestretch}{1}\selectfont
\centering
\caption{The journey by formation redshift}
\label{tab:results}
\footnotesize\setlength{\tabcolsep}{7pt}
\renewcommand{\arraystretch}{1.3}
\begin{tabular}{cIcIcIcIcIcIcIc}
\toprule
$z_{\rm src}$ & $t$ & $N_{\rm cat}$ & $n_{\rm LRD}$ & $\rho_{\rm org}$ & $N_{\rm hab}$ & \multicolumn{2}{c}{$N_{\rm life}/\Pabi$} \\
\cmidrule(lr){7-8}
 & Gyr & & cMpc$^{-3}$ & $\Msun\,\cGpc^{-3}$ & & median & 16--84\% \\
\midrule
% Rows generated by src/lrd_life_pathway_model.py (data/table_rows_main.tex); the constant dose columns are dropped.
2.25 & 2.91 & 2 & $1.9\times10^{-6}$ & $1.0\times10^{2}$ & $1.1\times10^{11}$ & $4.1\times10^{8}$ & $8.5\times10^{6}$--$6.3\times10^{9}$ \\
2.75 & 2.36 & 2 & $1.9\times10^{-6}$ & $1.3\times10^{2}$ & $9.6\times10^{10}$ & $3.5\times10^{8}$ & $7.2\times10^{6}$--$5.3\times10^{9}$ \\
3.25 & 1.96 & 5 & $5.0\times10^{-6}$ & $7.3\times10^{2}$ & $4.1\times10^{11}$ & $1.5\times10^{9}$ & $3.0\times10^{7}$--$2.3\times10^{10}$ \\
3.75 & 1.66 & 8 & $8.3\times10^{-6}$ & $6.6\times10^{2}$ & $2.8\times10^{11}$ & $1.0\times10^{9}$ & $2.1\times10^{7}$--$1.6\times10^{10}$ \\
4.25 & 1.43 & 13 & $1.4\times10^{-5}$ & $5.4\times10^{3}$ & $1.8\times10^{12}$ & $6.5\times10^{9}$ & $1.3\times10^{8}$--$9.9\times10^{10}$ \\
4.75 & 1.25 & 36 & $4.2\times10^{-5}$ & $2.0\times10^{4}$ & $5.4\times10^{12}$ & $2.0\times10^{10}$ & $4.4\times10^{8}$--$2.9\times10^{11}$ \\
5.25 & 1.10 & 64 & $7.8\times10^{-5}$ & $1.7\times10^{4}$ & $3.6\times10^{12}$ & $1.3\times10^{10}$ & $2.7\times10^{8}$--$2.0\times10^{11}$ \\
5.75 & 0.98 & 32 & $4.4\times10^{-5}$ & $1.2\times10^{4}$ & $2.0\times10^{12}$ & $7.5\times10^{9}$ & $1.5\times10^{8}$--$1.1\times10^{11}$ \\
6.25 & 0.88 & 44 & $6.2\times10^{-5}$ & $1.9\times10^{4}$ & $2.8\times10^{12}$ & $1.0\times10^{10}$ & $2.1\times10^{8}$--$1.6\times10^{11}$ \\
6.75 & 0.80 & 25 & $3.6\times10^{-5}$ & $1.4\times10^{4}$ & $1.7\times10^{12}$ & $6.4\times10^{9}$ & $1.3\times10^{8}$--$9.7\times10^{10}$ \\
7.25 & 0.73 & 49 & $8.8\times10^{-5}$ & $2.6\times10^{4}$ & $2.8\times10^{12}$ & $1.0\times10^{10}$ & $2.1\times10^{8}$--$1.6\times10^{11}$ \\
7.75 & 0.67 & 15 & $2.4\times10^{-5}$ & $6.0\times10^{3}$ & $5.5\times10^{11}$ & $2.0\times10^{9}$ & $4.1\times10^{7}$--$3.1\times10^{10}$ \\
8.25 & 0.61 & 11 & $2.0\times10^{-5}$ & $5.3\times10^{3}$ & $4.3\times10^{11}$ & $1.6\times10^{9}$ & $3.2\times10^{7}$--$2.4\times10^{10}$ \\
8.75 & 0.57 & 9 & $2.0\times10^{-5}$ & $1.3\times10^{4}$ & $9.5\times10^{11}$ & $3.5\times10^{9}$ & $7.2\times10^{7}$--$5.2\times10^{10}$ \\
9.25 & 0.52 & 10 & $2.2\times10^{-5}$ & $6.9\times10^{3}$ & $4.3\times10^{11}$ & $1.6\times10^{9}$ & $3.2\times10^{7}$--$2.4\times10^{10}$ \\
9.75 & 0.49 & 3 & $7.2\times10^{-6}$ & $1.1\times10^{3}$ & $6.2\times10^{10}$ & $2.3\times10^{8}$ & $4.6\times10^{6}$--$3.4\times10^{9}$ \\
10.25 & 0.46 & 5 & $1.0\times10^{-5}$ & $4.4\times10^{3}$ & $2.2\times10^{11}$ & $8.0\times10^{8}$ & $1.6\times10^{7}$--$1.2\times10^{10}$ \\
10.75 & 0.43 & 4 & $1.1\times10^{-5}$ & $2.0\times10^{3}$ & $9.1\times10^{10}$ & $3.3\times10^{8}$ & $6.8\times10^{6}$--$5.1\times10^{9}$ \\
11.25 & 0.40 & 3 & $8.3\times10^{-6}$ & $1.7\times10^{3}$ & $7.0\times10^{10}$ & $2.6\times10^{8}$ & $5.2\times10^{6}$--$3.9\times10^{9}$ \\
11.75 & 0.38 & 1 & $2.3\times10^{-6}$ & $5.9\times10^{2}$ & $2.1\times10^{10}$ & $7.9\times10^{7}$ & $1.6\times10^{6}$--$1.2\times10^{9}$ \\
\bottomrule
\end{tabular}

\medskip
\begin{flushleft}\footnotesize Bins of width $\Delta z=0.5$ in the formation redshift of the planets, $z_{\rm src}$, that of the LRD whose gas forms them; $t$ is the cosmic age, and life emerges on these planets 0.1--1~Gyr later.  $N_{\rm hab}$ is the number of planets formed from LRD gas that stay habitable, and $N_{\rm life}/\Pabi$ the number able to sustain LRD-fed life, per bin in a comoving volume equal to the observable Universe.  The median dose ratios (Equation~\ref{eq:dose}) are the same in every bin: $D=1.1\times10^{-8}$ for recycled organics and $D=0.038$ for in-situ organics, of which a median fraction $f_{\rm pass}=0.04$ of the planets reach $D\geq1$.\end{flushleft}
\end{table}

\section{What the Earth inherited: estimates}
\label{sec:earth_appendix}

The Sun formed 4.57~Gyr ago at 5--7~kpc from the Galactic Centre and now lies at 8.2~kpc (Section~\ref{sec:milkyway}).  LRD chemistry can have reached the Earth through four connections, which we estimate here by extending our transport step with processes known from the Milky Way and the Solar System.  The first requires an LRD active at the Sun's birth; the second and third apply if the Galaxy was an LRD or absorbed one; and the fourth applies to the Galaxy's present core.

% placed here so that it follows Table~\ref{tab:results} in the one-column layout

The first connection is direct: the Sun could have formed in the gas of an LRD.  LRDs and close analogues are found down to $z\approx0.1$~\cite{2025ApJ...980L..34L,2026ApJ...997..364L,2026MNRAS.545f2235J,2026arXiv260514233P,2026AandA...709L..11D}, including an LRD-like galaxy at $z=0.433$~\cite{2026ApJ...999...30C}, the epoch of the Sun's birth.  Their abundance at $z=0.2$--$0.9$ is at least $7.5\times10^{-10}\,\cMpc^{-3}$~\cite{2026arXiv260521574L}, and at cosmic noon ($z\approx2$) it is $3.4\times10^{-5}\,\cMpc^{-3}$~\cite{2026arXiv260630253L}.  With the $0.1$--$0.9\,\Msun\,{\rm yr^{-1}}$ (median 0.3) of star formation in the gas of an average LRD in our model and a cosmic star-formation rate density of $0.024\,\Msun\,{\rm yr^{-1}\,Mpc^{-3}}$ at $z=0.42$~\cite{2014ARAandA..52..415M}, a star born at the Sun's epoch formed in LRD gas with probability $10^{-8}$ for the lower limit on the abundance, and up to $5\times10^{-4}$ if LRDs were as common then as at cosmic noon.  The Sun's birth in the Galactic disk lowers this probability for the Solar System, but does not exclude it.

The second connection is through the elements.  Stars formed in LRD gas make up $(0.3$--$1.8)\times10^{-4}$ of the stellar mass of a Milky-Way-mass galaxy, and the LRD hosts, whose stellar masses are $\approx5\times10^{7}\,\Msun$ on average~\cite{2025ApJ...988..246M} and which we take as $10^{7.5}$--$10^{8.5}\,\Msun$, make up $4\times10^{-4}$--$4\times10^{-3}$.  Their supernovae and winds returned carbon, nitrogen, and oxygen to gas that was recycled through the Galaxy for billions of years~\cite{2016ApJ...824...57C,2017MNRAS.470.4698A,2019MNRAS.490.4786G}, or, for an accreted LRD, delivered them to the Galaxy through the merger.  If these products mixed with the rest of the Galaxy's nucleosynthesis, a fraction $10^{-4}$--$4\times10^{-3}$ of the Earth's carbon, nitrogen, and oxygen atoms passed through an LRD host.

The third connection is through molecules carried on dust.  Gas-phase and icy organics are destroyed within tens of Myr (Section~\ref{sec:travel}), but part of the interstellar organic inventory is refractory.  Ultraviolet irradiation of interstellar ice analogues leaves a residue that contains amino acids~\cite{2002Natur.416..403M,2002Natur.416..401B}.  Carbonaceous chondrites carry insoluble organic matter with the isotopic signatures of cold interstellar chemistry~\cite{2006Sci...312..727B,2017ChEG...77..227A} and all five canonical nucleobases~\cite{2022NatCo..13.2008O}, and much of the Solar System's water and cometary ice was inherited from the interstellar medium~\cite{2014Sci...345.1590C,2019ARAandA..57..113A}.  Solid carriers also survive transport.  Dust grains larger than $\sim0.01\,\mu$m survive galactic winds, 60\% of those of $0.1\,\mu$m~\cite{2024ApJ...974...81R}, and presolar silicon carbide grains in the Murchison meteorite spent from 4~Myr to $\sim3$~Gyr in the interstellar medium before the Solar System formed~\cite{2020PNAS..117.1884H}.  Interstellar dust is destroyed in supernova shocks on a timescale $\tau_{\rm dust}=0.4$~Gyr for carbonaceous grains in the classic estimate and 2--3~Gyr in revised calculations; we adopt 0.4--3~Gyr~\cite{1994ApJ...433..797J,2015ApJ...803....7S}.  Insoluble organic matter carries most of the organic carbon in chondrites~\cite{2017ChEG...77..227A}, but the prebiotic molecules proper are mostly volatile, so we give them a minor refractory share, $f_{\rm ref}=10^{-2}$--$10^{-1}$.  We let this fraction leave the LRD in outflows or with the merger, return with the efficiency $\epsilon_{\rm launch}f_{\rm rec}\approx0.1$ of Section~\ref{sec:travel}, and survive with probability $\exp(-\Delta t/\tau_{\rm dust})$.  The value of $f_{\rm ref}$ is not constrained by observations, and the result scales linearly with it.  Mixed into the $\sim4\times10^{10}\,\Msun$ of gas that the Galaxy turned into stars before the Sun formed~\cite{2016ARAandA..54..529B}, and delivered with the median efficiency $\epsilon_{\rm planet}=10^{-8}$ of other exogenous organics (Section~\ref{sec:life}), an in-situ LRD phase at $z\approx6$ ($\Delta t\approx8$~Gyr) gives the Earth $10^{-1}$--$5\times10^{7}$~kg of LRD-made organics, or $10^{24}$--$10^{33}$ molecules, and an LRD phase at $z\approx2$ in a galaxy accreted later gives $10^{1}$--$10^{8}$~kg.  These are $\lesssim10^{-7}$ of the $10^{15}$--$10^{17}$~kg of exogenous organics the Earth received.

The fourth connection runs through the Galactic Centre.  Ruffini and Wang~\cite{2026ApJ..1000L..11R} argued that the Central Molecular Zone is the present-day counterpart of an LRD core.  Whether or not it descends from an LRD phase of the Galaxy, the CMZ exports its chemistry.  The Galactic nuclear wind carries cold gas at $\lesssim0.1\,\Msun\,{\rm yr^{-1}}$~\cite{2018ApJ...855...33D}, including molecular clouds~\cite{2020Natur.584..364D}, and the Sun formed closer to the Galactic Centre than it is today.  Let a refractory fraction $f_{\rm ref}$ of the CMZ organics ride this wind at 0.01--0.1~$\Msun\,{\rm yr^{-1}}$, from a tenth of the present upper limit to the limit itself, over a dust lifetime and mix into the $\sim10^{9}\,\Msun$ of molecular gas in the inner Galaxy~\cite{2015ARAandA..53..583H}.  If local molecular clouds have the organic fraction of the CMZ, the Galactic Centre then supplies $10^{-5}$--$3\times10^{-2}$ of the interstellar organics that the Solar System inherited; a lower local fraction raises this share.  This contribution is testable.  Gas in the Galactic Centre has $^{12}$C/$^{13}$C$\,\approx20$--$25$, compared with 89 in the Solar System~\cite{1994ARAandA..32..191W}, and D/H$\,=1.7\times10^{-6}$, nine times below the local interstellar value~\cite{2000Natur.405.1025L}.  Organic grains or individual compounds that combine a $^{13}$C enrichment with a low deuterium content, in the samples returned from the asteroids Ryugu and Bennu, which contain amino acids and all five nucleobases~\cite{2023NatCo..14.1292O,2025NatAs...9..199G}, would trace this inheritance.


\begin{thebibliography}{210}

\bibitem{2009ARAandA..47..427H}
E.~Herbst, E.~F.~van Dishoeck, Annu. Rev. Astron. Astrophys. {\bf 47}, 427 (2009).

\bibitem{2012AandARv..20...56C}
P.~Caselli, C.~Ceccarelli, Astron. and Astrophys. Rev. {\bf 20}, id.\,56 (2012).

\bibitem{2021PhR...893....1O}
K.~I.~Öberg, E.~A.~Bergin, Phys. Rep. {\bf 893}, 1 (2021).

\bibitem{2022ApJS..259...30M}
B.~A.~McGuire, Astrophys. J. Suppl. {\bf 259}, id.\,30 (2022).

\bibitem{1953Sci...117..528M}
S.~L.~Miller, Science {\bf 117}, 528 (1953).

\bibitem{2009Natur.459..239P}
M.~W.~Powner, B.~Gerland, J.~D.~Sutherland, Nature {\bf 459}, 239 (2009).

\bibitem{2015NatCh...7..301P}
B.~H.~Patel, C.~Percivalle, D.~J.~Ritson, C.~D.~Duffy, et al., Nature Chem. {\bf 7}, 301 (2015).

\bibitem{2024ApJ...963..129M}
J.~Matthee, R.~P.~Naidu, G.~Brammer, J.~Chisholm, et al., Astrophys. J. {\bf 963}, id.\,129 (2024).

\bibitem{2024ApJ...968...38K}
V.~Kokorev, K.~I.~Caputi, J.~E.~Greene, P.~Dayal, et al., Astrophys. J. {\bf 968}, id.\,38 (2024).

\bibitem{2025ApJ...986..126K}
D.~D.~Kocevski, S.~L.~Finkelstein, G.~Barro, A.~J.~Taylor, et al., Astrophys. J. {\bf 986}, id.\,126 (2025).

\bibitem{2022MNRAS.511.4490R}
F.~Rastegarnia, M.~T.~Mirtorabi, R.~Moradi, A.~Vafaei Sadr, et al., Mon. Not. R. Astron. Soc. {\bf 511}, 4490 (2022).

\bibitem{2024MNRAS.533.1976M}
R.~Moradi, F.~Rastegarnia, Y.~Wang, M.~T.~Mirtorabi, Mon. Not. R. Astron. Soc. {\bf 533}, 1976 (2024).

\bibitem{2025ApJS..276...37L}
Y.~Liu, J.~Li, L.~Gao, H.~Zhang, et al., Astrophys. J. Suppl. {\bf 276}, id.\,37 (2025).

\bibitem{2026ApJ...997..364L}
X.~Lin, X.~Fan, Z.~Cai, F.~Bian, et al., Astrophys. J. {\bf 997}, id.\,364 (2026).

\bibitem{2026arXiv260521574L}
X.~Lin, X.~Fan, Z.~Cai, Y.~Liu, et al., arXiv:2605.21574 (2026).

\bibitem{2026Natur.649..574R}
V.~Rusakov, D.~Watson, G.~P.~Nikopoulos, G.~Brammer, et al., Nature {\bf 649}, 574 (2026).

\bibitem{2026Natur.653.1017J}
I.~Juodžbalis, C.~Marconcini, F.~D'Eugenio, R.~Maiolino, et al., Nature {\bf 653}, 1017 (2026).

\bibitem{2026ApJ..1000L..11R}
R.~Ruffini, Y.~Wang, Astrophys. J. Lett. {\bf 1000}, id.\,L11 (2026).

\bibitem{1996ARAandA..34..645M}
M.~Morris, E.~Serabyn, Annu. Rev. Astron. Astrophys. {\bf 34}, 645 (1996).

\bibitem{2006AandA...455..971R}
M.~A.~Requena-Torres, J.~Martín-Pintado, A.~Rodríguez-Franco, S.~Martín, et al., Astron. and Astrophys. {\bf 455}, 971 (2006).

\bibitem{2008ApJ...672..352R}
M.~A.~Requena-Torres, J.~Martín-Pintado, S.~Martín, M.~R.~Morris, Astrophys. J. {\bf 672}, 352 (2008).

\bibitem{2018MNRAS.478.2962Z}
S.~Zeng, I.~Jiménez-Serra, V.~M.~Rivilla, S.~Martín, et al., Mon. Not. R. Astron. Soc. {\bf 478}, 2962 (2018).

\bibitem{2020AandA...641A...6P}
P.~Collaboration, N.~Aghanim, Y.~Akrami, M.~Ashdown, et al., Astron. and Astrophys. {\bf 641}, id.\,A6 (2020).

\bibitem{2003PASP..115..763C}
G.~Chabrier, Publ. Astron. Soc. Pacific {\bf 115}, 763 (2003).

\bibitem{2002AsBio...2..293L}
C.~H.~Lineweaver, T.~M.~Davis, Astrobiology {\bf 2}, 293 (2002).

\bibitem{2020PNAS..11711995K}
D.~Kipping, Proc. Natl. Acad. Sci. USA {\bf 117}, 11995 (2020).

\bibitem{2012PNAS..109..395S}
D.~S.~Spiegel, E.~L.~Turner, Proc. Natl. Acad. Sci. USA {\bf 109}, 395 (2012).

\bibitem{1968ApJ...151..393S}
M.~Schmidt, Astrophys. J. {\bf 151}, 393 (1968).

\bibitem{1980ApJ...235..694A}
Y.~Avni, J.~N.~Bahcall, Astrophys. J. {\bf 235}, 694 (1980).

\bibitem{2025ApJ...980L..27I}
K.~Inayoshi, R.~Maiolino, Astrophys. J. Lett. {\bf 980}, id.\,L27 (2025).

\bibitem{2025AandA...701A.168D}
A.~de Graaff, H.-W.~Rix, R.~P.~Naidu, I.~Labbé, et al., Astron. and Astrophys. {\bf 701}, id.\,A168 (2025).

\bibitem{2026Natur.656..329N}
R.~P.~Naidu, J.~Matthee, H.~Katz, A.~de Graaff, et al., Nature {\bf 656}, 329 (2026).

\bibitem{2025MNRAS.544.3407K}
D.~Kido, K.~Ioka, K.~Hotokezaka, K.~Inayoshi, et al., Mon. Not. R. Astron. Soc. {\bf 544}, 3407 (2025).

\bibitem{2025ApJ...991L..10S}
D.~J.~Setton, J.~E.~Greene, J.~S.~Spilker, C.~C.~Williams, et al., Astrophys. J. Lett. {\bf 991}, id.\,L10 (2025).

\bibitem{2025AandA...700A.231X}
M.~Xiao, P.~A.~Oesch, L.~Bing, D.~Elbaz, et al., Astron. and Astrophys. {\bf 700}, id.\,A231 (2025).

\bibitem{2026ApJ...997..218A}
H.~B.~Akins, C.~M.~Casey, J.~Chisholm, D.~A.~Berg, et al., Astrophys. J. {\bf 997}, id.\,218 (2026).

\bibitem{2025ApJ...990L..61C}
C.~M.~Casey, H.~B.~Akins, S.~L.~Finkelstein, M.~Franco, et al., Astrophys. J. Lett. {\bf 990}, id.\,L61 (2025).

\bibitem{2014ARAandA..52..415M}
P.~Madau, M.~Dickinson, Annu. Rev. Astron. Astrophys. {\bf 52}, 415 (2014).

\bibitem{2026arXiv260631515N}
G.~P.~Nikopoulos, D.~Watson, C.~L.~Pollock, A.~Sneppen, et al., arXiv:2606.31515 (2026).

\bibitem{2026arXiv260909271K}
V.~Kokorev, J.~Chisholm, R.~P.~Naidu, M.~Gieles, et al., arXiv:2609.09271 (2026).

\bibitem{2023AandA...677A..88B}
A.~J.~Bunker, A.~Saxena, A.~J.~Cameron, C.~J.~Willott, et al., Astron. and Astrophys. {\bf 677}, id.\,A88 (2023).

\bibitem{2024AandA...684A..75C}
M.~Curti, R.~Maiolino, E.~Curtis-Lake, J.~Chevallard, et al., Astron. and Astrophys. {\bf 684}, id.\,A75 (2024).

\bibitem{2025NatAs...9..155Z}
J.~A.~Zavala, M.~Castellano, H.~B.~Akins, T.~J.~L.~C.~Bakx, et al., Nature Astron. {\bf 9}, 155 (2025).

\bibitem{2025ApJ...988...19S}
S.~Schouws, R.~J.~Bouwens, K.~Ormerod, R.~Smit, et al., Astrophys. J. {\bf 988}, id.\,19 (2025).

\bibitem{2026ApJ..1001...38T}
M.~Tang, D.~P.~Stark, C.~A.~Mason, V.~Gelli, et al., Astrophys. J. {\bf 1001}, id.\,38 (2026).

\bibitem{2026AandA...708A.203P}
C.~L.~Pollock, R.~Gottumukkala, K.~E.~Heintz, G.~B.~Brammer, et al., Astron. and Astrophys. {\bf 708}, id.\,A203 (2026).

\bibitem{2020AsBio..20.1048J}
I.~Jiménez-Serra, J.~Martín-Pintado, V.~M.~Rivilla, L.~Rodríguez-Almeida, et al., Astrobiology {\bf 20}, 1048 (2020).

\bibitem{2021PNAS..11801314R}
V.~M.~Rivilla, I.~Jiménez-Serra, J.~Martín-Pintado, C.~Briones, et al., Proc. Natl. Acad. Sci. USA {\bf 118}, id.\,e2101314118 (2021).

\bibitem{2018ApJ...853L..19S}
M.~Sewiło, R.~Indebetouw, S.~B.~Charnley, S.~Zahorecz, et al., Astrophys. J. Lett. {\bf 853}, id.\,L19 (2018).

\bibitem{2022ApJ...931..102S}
M.~Sewiło, M.~Cordiner, S.~B.~Charnley, J.~M.~Oliveira, et al., Astrophys. J. {\bf 931}, id.\,102 (2022).

\bibitem{2016ApJ...827...72S}
T.~Shimonishi, T.~Onaka, A.~Kawamura, Y.~Aikawa, Astrophys. J. {\bf 827}, id.\,72 (2016).

\bibitem{2020ApJ...891..164S}
T.~Shimonishi, A.~Das, N.~Sakai, K.~E.~I.~Tanaka, et al., Astrophys. J. {\bf 891}, id.\,164 (2020).

\bibitem{2023ApJ...946L..41S}
T.~Shimonishi, K.~E.~I.~Tanaka, Y.~Zhang, K.~Furuya, Astrophys. J. Lett. {\bf 946}, id.\,L41 (2023).

\bibitem{2023MNRAS.523.3516C}
A.~J.~Cameron, H.~Katz, M.~P.~Rey, A.~Saxena, Mon. Not. R. Astron. Soc. {\bf 523}, 3516 (2023).

\bibitem{2024ApJ...966...92S}
P.~Senchyna, A.~Plat, D.~P.~Stark, G.~C.~Rudie, et al., Astrophys. J. {\bf 966}, id.\,92 (2024).

\bibitem{2024ApJ...972..143C}
M.~Castellano, L.~Napolitano, A.~Fontana, G.~Roberts-Borsani, et al., Astrophys. J. {\bf 972}, id.\,143 (2024).

\bibitem{2026OJAp....956033N}
R.~P.~Naidu, P.~A.~Oesch, G.~Brammer, A.~Weibel, et al., Open J. Astrophys. {\bf 9}, 56033 (2026).

\bibitem{2026MNRAS.tmp.1597R}
V.~Rusakov, C.~J.~Conselice, T.~Harvey, J.~C.~J.~D'Silva, et al., Mon. Not. R. Astron. Soc., in press; arXiv:2606.18334 (2026).

\bibitem{2025ApJ...994L..29Z}
P.~Zhu, L.~J.~Kewley, T.~Y.-Y.~Hsiao, J.~Trussler, Astrophys. J. Lett. {\bf 994}, id.\,L29 (2025).

\bibitem{2024AandA...689A.152D}
F.~D'Eugenio, R.~Maiolino, S.~Carniani, J.~Chevallard, et al., Astron. and Astrophys. {\bf 689}, id.\,A152 (2024).

\bibitem{2014ApJ...797...69R}
I.~U.~Roederer, H.~R.~Jacobson, T.~Thanathibodee, A.~Frebel, et al., Astrophys. J. {\bf 797}, id.\,69 (2014).

\bibitem{2014ApJ...796L..24J}
H.~R.~Jacobson, T.~Thanathibodee, A.~Frebel, I.~U.~Roederer, et al., Astrophys. J. Lett. {\bf 796}, id.\,L24 (2014).

\bibitem{2017AandA...602A.105H}
A.~N.~Heays, A.~D.~Bosman, E.~F.~van Dishoeck, Astron. and Astrophys. {\bf 602}, id.\,A105 (2017).

\bibitem{2001ApJ...550L..95E}
P.~Ehrenfreund, M.~P.~Bernstein, J.~P.~Dworkin, S.~A.~Sandford, et al., Astrophys. J. Lett. {\bf 550}, id.\,L95 (2001).

\bibitem{2017ApJ...835..265W}
F.~Walter, A.~D.~Bolatto, A.~K.~Leroy, S.~Veilleux, et al., Astrophys. J. {\bf 835}, id.\,265 (2017).

\bibitem{2013Natur.499..450B}
A.~D.~Bolatto, S.~R.~Warren, A.~K.~Leroy, F.~Walter, et al., Nature {\bf 499}, 450 (2013).

\bibitem{2019ApJ...881...43K}
N.~Krieger, A.~D.~Bolatto, F.~Walter, A.~K.~Leroy, et al., Astrophys. J. {\bf 881}, id.\,43 (2019).

\bibitem{2019MNRAS.483.4586F}
A.~Fluetsch, R.~Maiolino, S.~Carniani, A.~Marconi, et al., Mon. Not. R. Astron. Soc. {\bf 483}, 4586 (2019).

\bibitem{2016ApJ...824...57C}
C.~R.~Christensen, R.~Davé, F.~Governato, A.~Pontzen, et al., Astrophys. J. {\bf 824}, id.\,57 (2016).

\bibitem{2017MNRAS.470.4698A}
D.~Anglés-Alcázar, C.-A.~Faucher-Giguère, D.~Kereš, P.~F.~Hopkins, et al., Mon. Not. R. Astron. Soc. {\bf 470}, 4698 (2017).

\bibitem{2019MNRAS.490.4786G}
R.~J.~J.~Grand, F.~van de Voort, J.~Zjupa, F.~Fragkoudi, et al., Mon. Not. R. Astron. Soc. {\bf 490}, 4786 (2019).

\bibitem{2015ApJ...805..158S}
E.~Scannapieco, M.~Brüggen, Astrophys. J. {\bf 805}, id.\,158 (2015).

\bibitem{2018MNRAS.480L.111G}
M.~Gronke, S.~P.~Oh, Mon. Not. R. Astron. Soc. {\bf 480}, id.\,L111 (2018).

\bibitem{2023MNRAS.524.1258N}
K.~A.~Noon, M.~R.~Krumholz, E.~M.~Di Teodoro, N.~M.~McClure-Griffiths, et al., Mon. Not. R. Astron. Soc. {\bf 524}, 1258 (2023).

\bibitem{2024ApJ...974...81R}
H.~M.~Richie, E.~E.~Schneider, M.~W.~Abruzzo, P.~Torrey, Astrophys. J. {\bf 974}, id.\,81 (2024).

\bibitem{2025AandA...698A.261B}
M.~Bouvier, S.~Viti, J.~G.~Mangum, C.~Eibensteiner, et al., Astron. and Astrophys. {\bf 698}, id.\,A261 (2025).

\bibitem{2025MNRAS.540..774D}
Q.~Duan, C.~J.~Conselice, Q.~Li, D.~Austin, et al., Mon. Not. R. Astron. Soc. {\bf 540}, 774 (2025).

\bibitem{2025MNRAS.540.2146P}
D.~Puskás, S.~Tacchella, C.~Simmonds, K.~Hainline, et al., Mon. Not. R. Astron. Soc. {\bf 540}, 2146 (2025).

\bibitem{2026arXiv260218068C}
A.~Calabrò, L.~Pentericci, M.~Llerena, S.~Rossi, et al., arXiv:2602.18068 (2026).

\bibitem{2024arXiv241214246T}
T.~S.~Tanaka, J.~D.~Silverman, K.~Shimasaku, J.~Arita, et al., arXiv:2412.14246 (2024).

\bibitem{2026ApJ..1002L...4B}
J.~F.~W.~Baggen, M.~T.~Scoggins, P.~van Dokkum, Z.~Haiman, et al., Astrophys. J. Lett. {\bf 1002}, id.\,L4 (2026).

\bibitem{2026arXiv260527903B}
A.~J.~Barger, L.~L.~Cowie, arXiv:2605.27903 (2026).

\bibitem{2025ApJ...989L..50C}
M.~Carranza-Escudero, C.~J.~Conselice, N.~Adams, T.~Harvey, et al., Astrophys. J. Lett. {\bf 989}, id.\,L50 (2025).

\bibitem{2022ApJ...940..168C}
C.~J.~Conselice, C.~J.~Mundy, L.~Ferreira, K.~Duncan, Astrophys. J. {\bf 940}, id.\,168 (2022).

\bibitem{1996ApJ...471..115B}
J.~E.~Barnes, L.~Hernquist, Astrophys. J. {\bf 471}, 115 (1996).

\bibitem{2009ApJ...691.1168H}
P.~F.~Hopkins, T.~J.~Cox, J.~D.~Younger, L.~Hernquist, Astrophys. J. {\bf 691}, 1168 (2009).

\bibitem{2001AandA...378...51B}
J.~Braine, P.-A.~Duc, U.~Lisenfeld, V.~Charmandaris, et al., Astron. and Astrophys. {\bf 378}, 51 (2001).

\bibitem{2013LNP...861..327D}
P.-A.~Duc, F.~Renaud, Lecture Notes in Physics {\bf 861}, 327 (2013).

\bibitem{2001AandA...372L..29A}
S.~Aalto, S.~Hüttemeister, A.~G.~Polatidis, Astron. and Astrophys. {\bf 372}, id.\,L29 (2001).

\bibitem{2007MNRAS.374.1227B}
M.~Boylan-Kolchin, C.-P.~Ma, Mon. Not. R. Astron. Soc. {\bf 374}, 1227 (2007).

\bibitem{2023ApJ...955L..12B}
J.~F.~W.~Baggen, P.~van Dokkum, I.~Labbé, G.~Brammer, et al., Astrophys. J. Lett. {\bf 955}, id.\,L12 (2023).

\bibitem{2017ApJ...839..114J}
P.~Jáchym, M.~Sun, J.~D.~P.~Kenney, L.~Cortese, et al., Astrophys. J. {\bf 839}, id.\,114 (2017).

\bibitem{2026ApJ..1000...90I}
K.~Inayoshi, K.~Murase, K.~Kashiyama, Astrophys. J. {\bf 1000}, id.\,90 (2026).

\bibitem{2026ApJ...998L...4S}
A.~D.~Santarelli, E.~Farag, E.~P.~Bellinger, P.~Natarajan, et al., Astrophys. J. Lett. {\bf 998}, id.\,L4 (2026).

\bibitem{2018MNRAS.475.3511G}
M.~Y.~Grudić, P.~F.~Hopkins, C.-A.~Faucher-Giguère, E.~Quataert, et al., Mon. Not. R. Astron. Soc. {\bf 475}, 3511 (2018).

\bibitem{2013MNRAS.429..987L}
S.~N.~Longmore, J.~Bally, L.~Testi, C.~R.~Purcell, et al., Mon. Not. R. Astron. Soc. {\bf 429}, 987 (2013).

\bibitem{2020MNRAS.493.2872C}
M.~Chevance, J.~M.~D.~Kruijssen, A.~P.~S.~Hygate, A.~Schruba, et al., Mon. Not. R. Astron. Soc. {\bf 493}, 2872 (2020).

\bibitem{2012ApJ...758...48Y}
C.-C.~Yang, M.~Krumholz, Astrophys. J. {\bf 758}, id.\,48 (2012).

\bibitem{2018MNRAS.475.2236K}
M.~R.~Krumholz, Y.-S.~Ting, Mon. Not. R. Astron. Soc. {\bf 475}, 2236 (2018).

\bibitem{2021AJ....161...36B}
S.~Bryson, M.~Kunimoto, R.~K.~Kopparapu, J.~L.~Coughlin, et al., Astron. J. {\bf 161}, id.\,36 (2021).

\bibitem{2015ApJ...807...45D}
C.~D.~Dressing, D.~Charbonneau, Astrophys. J. {\bf 807}, id.\,45 (2015).

\bibitem{2020MNRAS.498.2249H}
D.~C.~Hsu, E.~B.~Ford, R.~Terrien, Mon. Not. R. Astron. Soc. {\bf 498}, 2249 (2020).

\bibitem{2025ApJ...986L...3P}
E.~K.~Pass, D.~Charbonneau, A.~Vanderburg, Astrophys. J. Lett. {\bf 986}, id.\,L3 (2025).

\bibitem{2016PhR...663....1S}
A.~L.~Shields, S.~Ballard, J.~A.~Johnson, Phys. Rep. {\bf 663}, 1 (2016).

\bibitem{2018JCAP...05..020L}
M.~Lingam, A.~Loeb, J. Cosmol. Astropart. Phys. {\bf 2018}, id.\,020 (2018).

\bibitem{2012ApJ...751...81J}
J.~L.~Johnson, H.~Li, Astrophys. J. {\bf 751}, id.\,81 (2012).

\bibitem{2018AJ....155...89P}
E.~A.~Petigura, G.~W.~Marcy, J.~N.~Winn, L.~M.~Weiss, et al., Astron. J. {\bf 155}, id.\,89 (2018).

\bibitem{2024AJ....168..128B}
K.~M.~Boley, J.~L.~Christiansen, J.~Zink, K.~Hardegree-Ullman, et al., Astron. J. {\bf 168}, id.\,128 (2024).

\bibitem{2020MNRAS.491..903W}
A.~J.~Winter, J.~M.~D.~Kruijssen, M.~Chevance, B.~W.~Keller, et al., Mon. Not. R. Astron. Soc. {\bf 491}, 903 (2020).

\bibitem{2021MNRAS.501.1782C}
F.~Concha-Ramírez, M.~J.~C.~Wilhelm, S.~Portegies Zwart, S.~E.~van Terwisga, et al., Mon. Not. R. Astron. Soc. {\bf 501}, 1782 (2021).

\bibitem{2013MNRAS.431...63T}
T.~A.~Thompson, Mon. Not. R. Astron. Soc. {\bf 431}, 63 (2013).

\bibitem{2016ApJ...827...54D}
R.~Di Stefano, A.~Ray, Astrophys. J. {\bf 827}, id.\,54 (2016).

\bibitem{2015PNAS..11214518B}
E.~A.~Bell, P.~Boehnke, T.~M.~Harrison, W.~L.~Mao, Proc. Natl. Acad. Sci. USA {\bf 112}, 14518 (2015).

\bibitem{2024NatEE...8.1654M}
E.~R.~R.~Moody, S.~Álvarez-Carretero, T.~A.~Mahendrarajah, J.~W.~Clark, et al., Nature Ecol. Evol. {\bf 8}, 1654 (2024).

\bibitem{2003ApJ...585.1169G}
N.~Gehrels, C.~M.~Laird, C.~H.~Jackman, J.~K.~Cannizzo, et al., Astrophys. J. {\bf 585}, 1169 (2003).

\bibitem{2014PhRvL.113w1102P}
T.~Piran, R.~Jimenez, Phys. Rev. Lett. {\bf 113}, id.\,231102 (2014).

\bibitem{2015ApJ...810...41L}
Y.~Li, B.~Zhang, Astrophys. J. {\bf 810}, id.\,41 (2015).

\bibitem{2026ApJ...998..171Z}
S.-R.~Zhang, Y.~Wang, Y.-F.~Yuan, H.~Tagawa, et al., Astrophys. J. {\bf 998}, id.\,171 (2026).

\bibitem{2026JHEAp..5200576W}
Y.~Wang, R.~Ruffini, J. High Energy Astrophys. {\bf 52}, id.\,100576 (2026).

\bibitem{2018MNRAS.479..171F}
J.~C.~Forbes, A.~Loeb, Mon. Not. R. Astron. Soc. {\bf 479}, 171 (2018).

\bibitem{2017NatSR...716626B}
A.~Balbi, F.~Tombesi, Sci. Rep. {\bf 7}, id.\,16626 (2017).

\bibitem{2024ApJ...974L..26Y}
M.~Yue, A.-C.~Eilers, T.~T.~Ananna, C.~Panagiotou, et al., Astrophys. J. Lett. {\bf 974}, id.\,L26 (2024).

\bibitem{1966Sci...153...72S}
R.~Sanchez, J.~Ferris, L.~E.~Orgel, Science {\bf 153}, 72 (1966).

\bibitem{2017PNAS..11411327P}
B.~K.~D.~Pearce, R.~E.~Pudritz, D.~A.~Semenov, T.~K.~Henning, Proc. Natl. Acad. Sci. USA {\bf 114}, 11327 (2017).

\bibitem{2022ApJ...932....9P}
B.~K.~D.~Pearce, K.~Molaverdikhani, R.~E.~Pudritz, T.~Henning, et al., Astrophys. J. {\bf 932}, id.\,9 (2022).

\bibitem{1990Sci...249..366C}
C.~F.~Chyba, P.~J.~Thomas, L.~Brookshaw, C.~Sagan, Science {\bf 249}, 366 (1990).

\bibitem{1992Natur.355..125C}
C.~Chyba, C.~Sagan, Nature {\bf 355}, 125 (1992).

\bibitem{2019ESC.....3.1792R}
M.~Rubin, D.~V.~Bekaert, M.~W.~Broadley, M.~N.~Drozdovskaya, et al., ACS Earth Space Chem. {\bf 3}, 1792 (2019).

\bibitem{2019MNRAS.490...50D}
M.~N.~Drozdovskaya, E.~F.~van Dishoeck, M.~Rubin, J.~K.~Jørgensen, et al., Mon. Not. R. Astron. Soc. {\bf 490}, 50 (2019).

\bibitem{2021SciA....7.3632L}
J.~Li, E.~A.~Bergin, G.~A.~Blake, F.~J.~Ciesla, et al., Sci. Adv. {\bf 7}, id.\,eabd3632 (2021).

\bibitem{2005Sci...308.1014T}
F.~Tian, O.~B.~Toon, A.~A.~Pavlov, H.~De Sterck, Science {\bf 308}, 1014 (2005).

\bibitem{2018CSHPB..1034801J}
G.~F.~Joyce, J.~W.~Szostak, Cold Spring Harb. Perspect. Biol. {\bf 10}, id.\,a034801 (2018).

\bibitem{2016PNAS..113.8127S}
C.~Scharf, L.~Cronin, Proc. Natl. Acad. Sci. USA {\bf 113}, 8127 (2016).

\bibitem{1999astro.ph..5116H}
D.~W.~Hogg, arXiv:astro-ph/9905116 (1999).

\bibitem{2026AandA...711A..24E}
E.~Collaboration, L.~Bisigello, G.~Rodighiero, S.~Fotopoulou, et al., Astron. and Astrophys. {\bf 711}, id.\,A24 (2026).

\bibitem{2026arXiv260630253L}
F.~Loiacono, R.~Gilli, M.~Mignoli, M.~Brusa, et al., arXiv:2606.30253 (2026).

\bibitem{2026arXiv260407138R}
P.~Rinaldi, K.~Hainline, F.~D'Eugenio, P.~G.~Pérez-González, et al., arXiv:2604.07138 (2026).

\bibitem{2016AandA...592A..96G}
R.~Gobat, S.~E.~Hong, Astron. and Astrophys. {\bf 592}, id.\,A96 (2016).

\bibitem{2012MNRAS.421..621B}
I.~K.~Baldry, S.~P.~Driver, J.~Loveday, E.~N.~Taylor, et al., Mon. Not. R. Astron. Soc. {\bf 421}, 621 (2012).

\bibitem{2015ApJ...806...96L}
T.~C.~Licquia, J.~A.~Newman, Astrophys. J. {\bf 806}, id.\,96 (2015).

\bibitem{2016ARAandA..54..529B}
J.~Bland-Hawthorn, O.~Gerhard, Annu. Rev. Astron. Astrophys. {\bf 54}, 529 (2016).

\bibitem{2017MNRAS.465.2212S}
E.~Starkenburg, K.~A.~Oman, J.~F.~Navarro, R.~A.~Crain, et al., Mon. Not. R. Astron. Soc. {\bf 465}, 2212 (2017).

\bibitem{2018MNRAS.480..652E}
K.~El-Badry, J.~Bland-Hawthorn, A.~Wetzel, E.~Quataert, et al., Mon. Not. R. Astron. Soc. {\bf 480}, 652 (2018).

\bibitem{2022ApJ...941...45R}
H.-W.~Rix, V.~Chandra, R.~Andrae, A.~M.~Price-Whelan, et al., Astrophys. J. {\bf 941}, id.\,45 (2022).

\bibitem{2025MNRAS.537.3730H}
D.~Horta, R.~P.~Schiavon, Mon. Not. R. Astron. Soc. {\bf 537}, 3730 (2025).

\bibitem{2024ApJS..272...30H}
C.~K.~Harada, C.~D.~Dressing, S.~R.~Kane, B.~A.~Ardestani, Astrophys. J. Suppl. {\bf 272}, id.\,30 (2024).

\bibitem{2024ApJS..271...55K}
J.~D.~Kirkpatrick, F.~Marocco, C.~R.~Gelino, Y.~Raghu, et al., Astrophys. J. Suppl. {\bf 271}, id.\,55 (2024).

\bibitem{2020MNRAS.498.5720W}
B.~S.~Wlodarczyk-Sroka, M.~A.~Garrett, A.~P.~V.~Siemion, Mon. Not. R. Astron. Soc. {\bf 498}, 5720 (2020).

\bibitem{2024ApJ...970L...2C}
C.~Cadieux, R.~Doyon, R.~J.~MacDonald, M.~Turbet, et al., Astrophys. J. Lett. {\bf 970}, id.\,L2 (2024).

\bibitem{2025ApJ...983L..40M}
N.~Madhusudhan, S.~Constantinou, M.~Holmberg, S.~Sarkar, et al., Astrophys. J. Lett. {\bf 983}, id.\,L40 (2025).

\bibitem{2025AandA...700A.284L}
R.~Luque, C.~Piaulet-Ghorayeb, M.~Radica, Q.~Xue, et al., Astron. and Astrophys. {\bf 700}, id.\,A284 (2025).

\bibitem{2018AJ....156..260W}
J.~T.~Wright, S.~Kanodia, E.~Lubar, Astron. J. {\bf 156}, id.\,260 (2018).

\bibitem{2021AJ....162...33G}
V.~Gajjar, K.~I.~Perez, A.~P.~V.~Siemion, G.~Foster, et al., Astron. J. {\bf 162}, id.\,33 (2021).

\bibitem{2025ApJ...989L..19P}
F.~Pacucci, A.~Loeb, Astrophys. J. Lett. {\bf 989}, id.\,L19 (2025).

\bibitem{2008ApJ...689.1044G}
A.~M.~Ghez, S.~Salim, N.~N.~Weinberg, J.~R.~Lu, et al., Astrophys. J. {\bf 689}, 1044 (2008).

\bibitem{2023MNRAS.525.4456B}
V.~Belokurov, A.~Kravtsov, Mon. Not. R. Astron. Soc. {\bf 525}, 4456 (2023).

\bibitem{2025ApJ...994...94T}
V.~Y.~Y.~Tan, A.~Muzzin, G.~T.~E.~Sarrouh, J.~Antwi-Danso, et al., Astrophys. J. {\bf 994}, id.\,94 (2025).

\bibitem{2018Natur.563...85H}
A.~Helmi, C.~Babusiaux, H.~H.~Koppelman, D.~Massari, et al., Nature {\bf 563}, 85 (2018).

\bibitem{2018MNRAS.478..611B}
V.~Belokurov, D.~Erkal, N.~W.~Evans, S.~E.~Koposov, et al., Mon. Not. R. Astron. Soc. {\bf 478}, 611 (2018).

\bibitem{2020MNRAS.498.2472K}
J.~M.~D.~Kruijssen, J.~L.~Pfeffer, M.~Chevance, A.~Bonaca, et al., Mon. Not. R. Astron. Soc. {\bf 498}, 2472 (2020).

\bibitem{2021MNRAS.500.1385H}
D.~Horta, R.~P.~Schiavon, J.~T.~Mackereth, J.~Pfeffer, et al., Mon. Not. R. Astron. Soc. {\bf 500}, 1385 (2021).

\bibitem{2025ApJ...980L..34L}
R.~Lin, Z.-Y.~Zheng, C.~Jiang, F.-T.~Yuan, et al., Astrophys. J. Lett. {\bf 980}, id.\,L34 (2025).

\bibitem{2026MNRAS.545f2235J}
X.~Ji, F.~D'Eugenio, I.~Juodžbalis, D.~J.~Walton, et al., Mon. Not. R. Astron. Soc. {\bf 545}, id.\,staf2235 (2026).

\bibitem{2026arXiv260514233P}
K.~Park, A.~Torralba, J.~Matthee, S.~Mascia, et al., arXiv:2605.14233 (2026).

\bibitem{2026AandA...709L..11D}
W.~Ding, X.~Kong, W.-J.~Guo, H.~Zou, et al., Astron. and Astrophys. {\bf 709}, id.\,L11 (2026).

\bibitem{2026ApJ...999...30C}
X.~Chen, K.~Ichikawa, M.~Akiyama, K.~Inayoshi, et al., Astrophys. J. {\bf 999}, id.\,30 (2026).

\bibitem{2012Sci...338..651C}
J.~N.~Connelly, M.~Bizzarro, A.~N.~Krot, Å.~Nordlund, et al., Science {\bf 338}, 651 (2012).

\bibitem{1996AandA...314..438W}
R.~Wielen, B.~Fuchs, C.~Dettbarn, Astron. and Astrophys. {\bf 314}, 438 (1996).

\bibitem{2013AandA...558A...9M}
I.~Minchev, C.~Chiappini, M.~Martig, Astron. and Astrophys. {\bf 558}, id.\,A9 (2013).

\bibitem{2023MNRAS.526.6088B}
J.~Baba, T.~R.~Saitoh, T.~Tsujimoto, Mon. Not. R. Astron. Soc. {\bf 526}, 6088 (2023).

\bibitem{2010ARAandA..48...47A}
F.~C.~Adams, Annu. Rev. Astron. Astrophys. {\bf 48}, 47 (2010).

\bibitem{2019AandA...625L..10G}
G.~Collaboration, R.~Abuter, A.~Amorim, M.~Bauböck, et al., Astron. and Astrophys. {\bf 625}, id.\,L10 (2019).

\bibitem{1994ARAandA..32..191W}
T.~L.~Wilson, R.~Rood, Annu. Rev. Astron. Astrophys. {\bf 32}, 191 (1994).

\bibitem{2000Natur.405.1025L}
D.~A.~Lubowich, J.~M.~Pasachoff, T.~J.~Balonek, T.~J.~Millar, et al., Nature {\bf 405}, 1025 (2000).

\bibitem{2019ApJS..241....3P}
M.~T.~Penny, B.~S.~Gaudi, E.~Kerins, N.~J.~Rattenbury, et al., Astrophys. J. Suppl. {\bf 241}, id.\,3 (2019).

\bibitem{2002astro.ph.10394H}
D.~W.~Hogg, I.~K.~Baldry, M.~R.~Blanton, D.~J.~Eisenstein, arXiv:astro-ph/0210394 (2002).

\bibitem{2019Sci...366...76B}
S.~Becker, J.~Feldmann, S.~Wiedemann, H.~Okamura, et al., Science {\bf 366}, 76 (2019).

\bibitem{1996Natur.381...59F}
J.~P.~Ferris, A.~R.~Hill, R.~Liu, L.~E.~Orgel, Nature {\bf 381}, 59 (1996).

\bibitem{2008Natur.454..122M}
S.~S.~Mansy, J.~P.~Schrum, M.~Krishnamurthy, S.~Tobé, et al., Nature {\bf 454}, 122 (2008).

\bibitem{2020AsBio..20..429D}
B.~Damer, D.~Deamer, Astrobiology {\bf 20}, 429 (2020).

\bibitem{2018SciA....4.3302R}
P.~B.~Rimmer, J.~Xu, S.~J.~Thompson, E.~Gillen, et al., Sci. Adv. {\bf 4}, id.\,eaar3302 (2018).

\bibitem{2020AJ....159...60G}
M.~N.~Günther, Z.~Zhan, S.~Seager, P.~B.~Rimmer, et al., Astron. J. {\bf 159}, id.\,60 (2020).

\bibitem{2017ApJ...843..110R}
S.~Ranjan, R.~Wordsworth, D.~D.~Sasselov, Astrophys. J. {\bf 843}, id.\,110 (2017).

\bibitem{2021AsBio..21.1099R}
P.~B.~Rimmer, S.~J.~Thompson, J.~Xu, D.~A.~Russell, et al., Astrobiology {\bf 21}, 1099 (2021).

\bibitem{2014ApJ...781...27C}
N.~B.~Cowan, D.~S.~Abbot, Astrophys. J. {\bf 781}, id.\,27 (2014).

\bibitem{2015NatGe...8..177T}
F.~Tian, S.~Ida, Nature Geosci. {\bf 8}, 177 (2015).

\bibitem{2022NatAs...6.1296K}
T.~Kimura, M.~Ikoma, Nature Astron. {\bf 6}, 1296 (2022).

\bibitem{2017MNRAS.468.2803S}
F.~Simpson, Mon. Not. R. Astron. Soc. {\bf 468}, 2803 (2017).

\bibitem{2018Life....8...55B}
J.~L.~Bada, J.~Korenaga, Life {\bf 8}, id.\,55 (2018).

\bibitem{2021Life...11.1142K}
J.~Korenaga, Life {\bf 11}, id.\,1142 (2021).

\bibitem{2005AsBio...5..515P}
M.~A.~Pasek, D.~S.~Lauretta, Astrobiology {\bf 5}, 515 (2005).

\bibitem{2020PNAS..117..883T}
J.~D.~Toner, D.~C.~Catling, Proc. Natl. Acad. Sci. USA {\bf 117}, 883 (2020).

\bibitem{2020ApJ...900L..38H}
N.~R.~Hinkel, H.~E.~Hartnett, P.~A.~Young, Astrophys. J. Lett. {\bf 900}, id.\,L38 (2020).

\bibitem{2020NatGe..13..344R}
D.~J.~Ritson, S.~J.~Mojzsis, J.~D.~Sutherland, Nature Geosci. {\bf 13}, 344 (2020).

\bibitem{2025ApJ...988..246M}
J.~Matthee, R.~P.~Naidu, G.~Kotiwale, L.~J.~Furtak, et al., Astrophys. J. {\bf 988}, id.\,246 (2025).

\bibitem{2002Natur.416..403M}
G.~M.~Muñoz Caro, U.~J.~Meierhenrich, W.~A.~Schutte, B.~Barbier, et al., Nature {\bf 416}, 403 (2002).

\bibitem{2002Natur.416..401B}
M.~P.~Bernstein, J.~P.~Dworkin, S.~A.~Sandford, G.~W.~Cooper, et al., Nature {\bf 416}, 401 (2002).

\bibitem{2006Sci...312..727B}
H.~Busemann, A.~F.~Young, C.~M.~O'D. Alexander, P.~Hoppe, et al., Science {\bf 312}, 727 (2006).

\bibitem{2017ChEG...77..227A}
C.~M.~O.~'D.~Alexander, G.~D.~Cody, B.~T.~De Gregorio, L.~R.~Nittler, et al., Chem. Erde {\bf 77}, 227 (2017).

\bibitem{2022NatCo..13.2008O}
Y.~Oba, Y.~Takano, Y.~Furukawa, T.~Koga, et al., Nature Commun. {\bf 13}, id.\,2008 (2022).

\bibitem{2014Sci...345.1590C}
L.~I.~Cleeves, E.~A.~Bergin, C.~M.~O.~'D.~Alexander, F.~Du, et al., Science {\bf 345}, 1590 (2014).

\bibitem{2019ARAandA..57..113A}
K.~Altwegg, H.~Balsiger, S.~A.~Fuselier, Annu. Rev. Astron. Astrophys. {\bf 57}, 113 (2019).

\bibitem{2020PNAS..117.1884H}
P.~R.~Heck, J.~Greer, L.~Kööp, R.~Trappitsch, et al., Proc. Natl. Acad. Sci. USA {\bf 117}, 1884 (2020).

\bibitem{1994ApJ...433..797J}
A.~P.~Jones, A.~G.~G.~M.~Tielens, D.~J.~Hollenbach, C.~F.~McKee, Astrophys. J. {\bf 433}, 797 (1994).

\bibitem{2015ApJ...803....7S}
J.~D.~Slavin, E.~Dwek, A.~P.~Jones, Astrophys. J. {\bf 803}, id.\,7 (2015).

\bibitem{2018ApJ...855...33D}
E.~M.~Di Teodoro, N.~M.~McClure-Griffiths, F.~J.~Lockman, S.~R.~Denbo, et al., Astrophys. J. {\bf 855}, id.\,33 (2018).

\bibitem{2020Natur.584..364D}
E.~M.~Di Teodoro, N.~M.~McClure-Griffiths, F.~J.~Lockman, L.~Armillotta, Nature {\bf 584}, 364 (2020).

\bibitem{2015ARAandA..53..583H}
M.~Heyer, T.~M.~Dame, Annu. Rev. Astron. Astrophys. {\bf 53}, 583 (2015).

\bibitem{2023NatCo..14.1292O}
Y.~Oba, T.~Koga, Y.~Takano, N.~O.~Ogawa, et al., Nature Commun. {\bf 14}, id.\,1292 (2023).

\bibitem{2025NatAs...9..199G}
D.~P.~Glavin, J.~P.~Dworkin, C.~M.~O.~Alexander, J.~C.~Aponte, et al., Nature Astron. {\bf 9}, 199 (2025).

\end{thebibliography}
\end{document}